      \documentclass[useAMS,usenatbib]{mn2e}      
\usepackage{epsfig}

\def\spose#1{\hbox to 0pt{#1\hss}} 
\def\lta{\mathrel{\spose{\lower 3pt\hbox{$\mathchar"218$}}      
     \raise 2.0pt\hbox{$\mathchar"13C$}}}      
\def\gta{\mathrel{\spose{\lower 3pt\hbox{$\mathchar"218$}}      
     \raise 2.0pt\hbox{$\mathchar"13E$}}}     
   
\title[The Chemical Evolution of Globular Clusters]{The Chemical Evolution 
of Globular Clusters - II. Metals and Fluorine}
 
\author[S\'anchez-Bl\'azquez et~al.] {P. S\'anchez-Bl\'azquez$^{1}$,
       A. Marcolini$^2$,
       B~K. Gibson$^{2,3,4}$,
       A.~I. Karakas$^5$,\newauthor K. Pilkington$^{2,3,4}$ and F. Calura$^2$ \\
       $^1$Departamento de F\'isica Te\'orica, Universidad Aut\'onoma de
       Madrid, E28049, Cantoblanco, Madrid, Spain \\
       $^2$Jeremiah Horrocks Institute,
       University of Central Lancashire, Preston, PR1~2HE, UK \\
       $^{3}$Department of Astronomy \& Physics, Saint Mary's University,
       Halifax, Nova Scotia, B3H~3C3, Canada\\
       $^{4}$Monash Centre for Astrophysics, School of Mathematical Sciences,
       Monash University, Clayton, VIC, 3800, Australia\\
       $^5$Research School of Astronomy \& Astrophysics, Mt Stromlo
       Observatory, Weston Creek, ACT 2611, Australia \\
       }
    
\pagerange{\pageref{firstpage}--\pageref{lastpage}} 
\pubyear{2011} 
 
\begin{document} 
 
\maketitle

\label{firstpage} 
 
\begin{abstract} 
In the first paper in this series, we proposed a new framework in which 
to model the chemical evolution of globular clusters. This model, is 
predicated upon the assumption that clusters form within an interstellar 
medium enriched locally by the ejecta of a single Type~Ia supernova and 
varying numbers of asymptotic giant branch stars, superimposed on an 
ambient medium pre-enriched by low-metallicity Type~II supernovae.  
Paper~I was concerned with the application of this model to the observed 
abundances of several reactive elements and so-called non-metals for 
three classical intermediate-metallicity clusters, with the hallmark of 
the work being the successful recovery of many of their well-known 
elemental and isotopic abundance anomalies. Here, we expand upon our 
initial analysis by (a) applying the model to a much broader range of 
metallicities (from the factor of three explored in Paper~I, to now, a 
factor of $\sim$50; i.e., essentially, the full range of Galactic 
globular cluster abundances, and (b) incorporating a broader suite of 
chemical species, including a number of iron-peak isotopes, heavier 
$\alpha$-elements, and fluorine. While allowing for an appropriate fine tuning of the 
model input parameters, most empirical globular cluster 
abundance trends are reproduced, 
our model would suggest the need for a 
higher production of calcium, silicon, and copper in low-metallicity (or 
so-called ``prompt'') Type~Ia supernovae than predicted in current 
stellar models in order to reproduce the observed trends in NGC~6752, 
and a factor of two reduction in carbon production from asymptotic giant 
branch stars to explain the observed trends between carbon and nitrogen. 
Observations of heavy-element isotopes produced primarily by Type~Ia 
supernovae, including those of titanium, iron, and nickel, could 
support/refute unequivocally our proposed framework, 
although currently the feasibility of the proposed observations is well beyond
current instrumental capabilities. Hydrodynamical
simulations would be necessary to study its viability from a dynamical 
point of view.
\end{abstract} 


\section{Introduction}  
\label{sec:introduction}  
 
Globular clusters (GCs) are gravitationally-bound systems containing 
hundreds of thousands of stars, each cluster orbiting about a parent 
galaxy.  The Milky Way (MW) itself has more than 150 associated GCs 
(Harris 1996). While conceptually simple, internally coeval systems, 
from a chemical perspective, GCs are extremely interesting in the sense 
that they show a large variation in the abundances of light elements 
(i.e, Li, C, N, O, F, Na, Mg, and Al) \citep[e.g.][and references 
therein]{smith1987, kraft1993, grundahl2002, cohen2002, yong2003, 
sneden2004, yong2005, pasquini2005, cohen2005,smith2005, carretta2006, 
gratton2007, bonifacio2007, marino2008} both internally to a given 
cluster, and between clusters. Conversely, the abundances of heavier 
$\alpha$ (e.g., Si, Ca, Ti), iron-peak (e.g., Fe, Ni, Cu), and 
neutron-capture elements (e.g., Ba, La, Eu) do not, in general, show the 
same star-to-star variation.

Correlated (and anti-correlated) elemental and isotopic (anomalous) 
trends between these various nuclei are observed from the main sequence 
turn-off through to the tip of the first ascent giant branch, and are 
not shared by the corresponding stars in the field \citep{gayandhi2009}. 
For this reason, the primary driver responsible for these anomalous 
abundance patterns is thought to be ``external'' and related with local 
environment, while ``internal'' mixing mechanisms are thought to be 
important only in causing the variations of C and N (and possibly O) in 
evolved giants. Such arguments led to the postulation of the 
``self-pollution'' scenario \citep{cottrell1981, dantona1983} as being 
perhaps the one most robust in explaining the chemical abundances 
anomalies in clusters \citep[see the review of][and references 
therein]{gratton2004}.  According to the self-pollution scenario, a 
previous generation of stars polluted the atmospheres of stars observed 
today or provided much of the material from which those stars formed 
\citep[e.g.][]{jehin1998, parmentier1999, tsujimoto2007}. Several models 
have been proposed with intermediate-mass asymptotic giant branch (AGB) 
stars \citep[e.g.,][]{cottrell1981, denissenkov2003, fenner2004, 
dantona2005, bekki2007} and winds from massive stars 
\citep[e.g.,][]{prantzos2006, decressin2007a, decressin2007b} as the 
most popular candidate polluters,  as they provide the most
simple explanation for the lack of
internal spread seen in Fe and Ca in most clusters.

Hydrodynamical simulations of GC formation and evolution 
under the ``classical'' scenario have been performed by D'Ercole et~al. (2008)\nocite{2008MNRAS.391..825D}
and Bekki (2010)\nocite{2010ApJ...724L..99B}, although some fine-tuning
to the stellar IMF of the second generation -- as well as the duration of 
the star formation -- were necessary to reproduce the chemical 
properties and the masses of the first and
second generation of stars and to prevent SN explosions in the second 
generation (D'Ercole et~al. 2008).

These classical scenarios still possess several problems that
require solution.
One of the main problems lies in understanding how a second
generation can form with a current total mass comparable to the first
generation, at least for the old MW GCs (see D'Ercole et~al. 2008\nocite{2008MNRAS.391..825D}
for a discussion of the problem) without invoking an
anomalous IMF (D'Antona \& Caloi 2004\nocite{2004ApJ...611..871D}; 
Decressin et~al. 2007\nocite{2007A&A...475..859D}) or 
extreme mass loss from the first 
generation stars 
during the GC evolution (by at least a factor of 10, e.g., 
D'Antona \& Caloi 2004; Bekki \& Norris 2006; Prantzos \& Charbonnel 2006;
D'Antona et~al. 2007; Decressin et~al. 2007). 
External pollution form stars in the field \citep{bekki2007} 
have been proposed to alleviate this problem.

Other, more speculative, models have considered, for example, variations 
in the shape of the initial mass function (IMF) \citep{smith1982, 
dantona2004, prantzos2006} or significant variations to the underlying 
stellar structure and associated nucleosynthesis (e.g., no hot bottom 
burning, no third dredge up, extra mixing, and/or overshooting).  Such 
models have certainly enjoyed success in explaining certain aspects of 
anomalous abundance patterns in GCs \citep{dantona2007,bekki2007}, but 
not in their entirety.

 In summary, although the "classical" scenario succeeds in reproducing 
many of the observed chemical abundance anomalies of GCs, it does still 
remain problematic, as does the lack of consensus concerning
the dominant formation processes of GCs.  As such, we considered
it valuable, and indeed still do, to consider alternate, potentially
viable, solutions.

In Paper~I \citep{marcolini2009}, we proposed a new, and somewhat 
unique, framework in which to model the chemical evolution of globular 
clusters.  
Contrary to previous ``self-pollution'' models, in our 
framework the first stars form with ``peculiar'' abundance patterns 
seeded by Population~III pre-enrichment, while the so-called ``normal'' 
stars form during a second phase of self-pollution from Population~II 
Type~II supernovae (SNeII). 
 Despite being restricted to an analytical chemical evolution
framework, and admittedly requiring confirmation via fully 
hydrodynamical simulations, 
we showed that our model successfully reproduced most of the 
well-known anti-correlations between various light elements and 
isotopes, while maintaining both constant iron {\it and} C+N+O 
abundances, and simultaneously recovering the empirical magnesium 
isotope patterns. The only major problem that the model encountered was 
in its underproduction of aluminum; in order to reproduce observations, 
a factor of 50 more Al production in intermediate-mass AGB stars was 
needed. We noted that such an underproduction could be accommodated 
within the stellar nucleosynthetic uncertainties, without compromising 
the predicted abundance pattern of Al in the Milky Way, via the use of 
Galactic chemical evolution modeling.

Our picture for globular cluster formation is predicated upon the 
assumption of localised pollution from a single Type~Ia and $\sim$100 
AGB stars.  Such conditions give a qualitative explanation for the 
complete absence of Galactic GCs with [Fe/H]$\le-2.4$ 
\citep[e.g.][]{harris1996}, while extremely metal-poor field stars 
([Fe/H]$\le -$3.0) exist in copious numbers \citep[e.g.][and references 
therein]{frebel2007}. 

In Paper~I, emphasis was placed on the evolution of the light elements 
within several intermediate-metallicity ([Fe/H]$\sim$$-$1.4) GCs 
(NGC~6752, M~13, and NGC~2808).  In Paper~II, we now extend the 
application of the model to both the more metal-poor (M~15 and NGC~6397, 
with [Fe/H]=$-$2.26 and $-$1.95, respectively - Harris 1996\nocite{harris1996}) and 
metal-rich (47~Tuc, with [Fe/H]=$-$0.76) regimes.  In light of the 
availability of recent observational work, we also explore the behaviour 
of fluorine in M~4, and the iron-peak elements (and isotopes) for 
NGC~6752 and 47~Tuc. In \S~\ref{sec:yields}, we discuss our adopted 
stellar yields, while in \S~\ref{sec:model}, the framework itself is 
summarised. Our results, both globally and on a cluster-by-cluster 
basis, are presented in \S~4$-$7.
  
\begin{figure}     
\begin{center}     
\psfig{figure=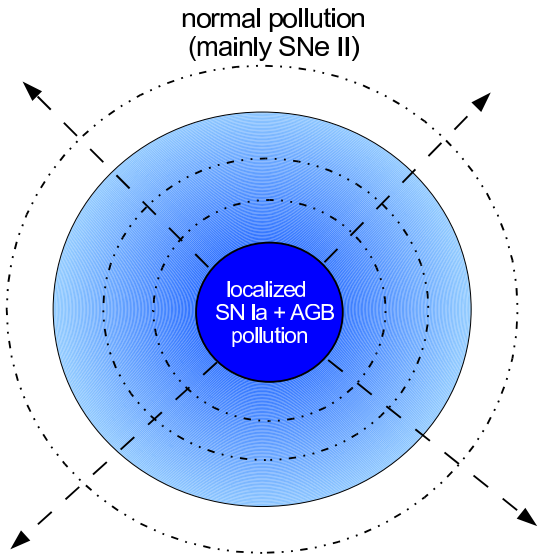,width=0.40\textwidth} 
\end{center}    
\caption{Schematic of the model outlined in Paper~I. Initially, a 
localised volume (inner blue region; solid line) pre-enriched with 
Population~III ejecta is polluted by a single SNIa and $\sim$100 
intermediate-mass AGB stars. After a new generation of stars is born, 
associated SNe~II begin to pollute and expand the inner volume, while 
mixing concurrently with the surrounding lower-[Fe/H] ISM.}
\label{fig:model}  
\end{figure} 
 
\section{Stellar Yields}
\label{sec:yields}

As discussed in Paper~I, we employ an IMF-weighted yield distribution 
for our chemical evolution predictions, based upon four sets of SNeII 
models, spanning the mass range 10-60 M$_\odot$ \citep{woosley1995, 
chieffi2004, kobayashi2006}, as summarised in Table~\ref{tab:sne_ii}.  
Therein, we only show the new elements analysed in this paper; we refer 
the reader to Tables 1 and 2 of Paper~I for the elements discussed in 
our earlier work. While in a global sense, the various yields are in 
reasonable agreement (particularly for the $\alpha$-elements), there are 
obviously those for which factors of $\sim$3 variations exist (e.g., 
several of the iron-peak elements, for which the sensitivity to the mass 
cut is most extreme). As in Paper~I, an average yield is derived by 
averaging between the compilations when agreement exists to within a 
factor of two, otherwise an admittedly arbitrary decision was made to 
adopt the compilation that best reproduces the abundances of 
low-metallicity ($-2.0 \le$[Fe/H]$\le -1.0$) halo stars 
\citep[e.g,][]{gratton2000}. The element with the largest variation is 
that of fluorine; this is due to the inclusion of the neutrino process 
by \citet{woosley1995} that increases by more than an order of magnitude 
the production of light elements (including $^7$Li and $^{19}$F) 
\citep[see][for further discussions]{timmes1995}. In this case we use 
the yields of \citet{woosley1995}. Unless stated differently, in the 
following we refer to the yields given in Table~\ref{tab:sne_ii} as the 
``Model'' yields.

\begin{table*}  
\centering 
\begin{minipage}{175mm}  
\caption{Mean SNeII stellar yields averaged over the progenitor mass 
range 10-60~M$_{\odot}$ for a Salpeter (1955) IMF for several 
compilations: W\&W=\citet{woosley1995}; C\&L=\citet{chieffi2004}; 
KOB=\citet{kobayashi2006}. The yields of Fe are given in solar masses 
while for different elements we show the [$X_{\rm i}$/Fe] ratio.}
\label{tab:sne_ii}    
\begin{tabular} {|l|c|c|c|c|c|c|c|c|c|c|c|c|c|}  
\hline \\   
SNe type &$\,$ Fe $\,$ & $\,$[F/Fe] $\,$ & $\,$ [Si/Fe] $\,$
&$\,$ [Ca/Fe] $\,$&$\,$ [Ti/Fe] $\,$ & $\,$ [V/Fe] $\,$ 
&$\,$ [Co/Fe] $\,$&$\,$ [Ni/Fe] $\,$ & $\,$ [Cu/Fe] \\  
 
\hline 
  
W\&W (Z=0.0002)  &   6.1e-2$^a$ & $-0.29$  & $+$0.45 & $+$0.31 & $-0.06$ & $-0.29$ & $-0.01$ & $+0.26$ &  $-$0.55 \\ 
W\&W (Z=0.002)   &   6.9e-2$^a$ & $+0.11$  & $+$0.42 & $+$0.28 & $-0.13$ & $-0.21$ & $+0.12$ & $+0.30$ &  $-$0.37 \\ 
C\&L (Z=0.0001)  &   1.0e-1     & $-2.69$  & $+$0.62 & $+$0.54 & $-0.22$ & $-0.31$ & $-0.39$ & $+0.34$ &  $-$0.55 \\ 
C\&L (Z=0.0001)  &   1.0e-1     & $-1.57$  & $+$0.60 & $+$0.51 & $-0.15$ & $-0.47$ & $-0.26$ & $+0.23$ &  $-$0.68 \\  
KOB  (Z=0.001)   &   7.5e-2     & $-1.08$  & $+$0.60 & $+$0.35 & $+0.03$ & $-0.31$ & $-0.35$ & $-0.35$ &  $-$0.39 \\ 
KOB  (Z=0.001+HN)&   1.1e-1     & $-1.20$  & $+$0.57 & $+$0.28 & $-0.13$ & $-0.25$ & $-0.17$ & $-0.21$ &  $-$0.41 \\ 
\hline  
Model (SNe II)            &   9.0e-2     &  $+$0.00  & $+$0.40 & $+0.30$ & $-0.05$ & $-0.30$ & $+0.00$ & $+$0.00 & $-$0.55 \\ 
\hline  
 
\end{tabular} 
\end{minipage} 
\end{table*}

\begin{figure*}     
\begin{center}     
\psfig{figure=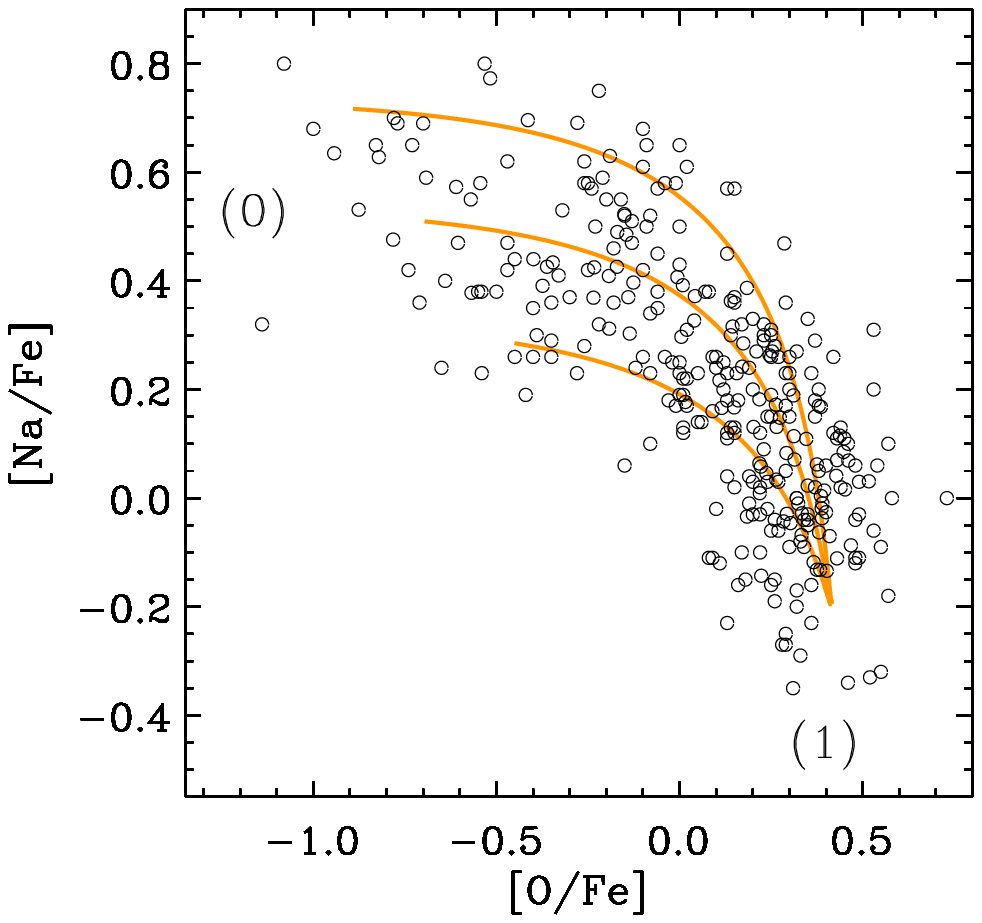,width=0.40\textwidth}
\psfig{figure=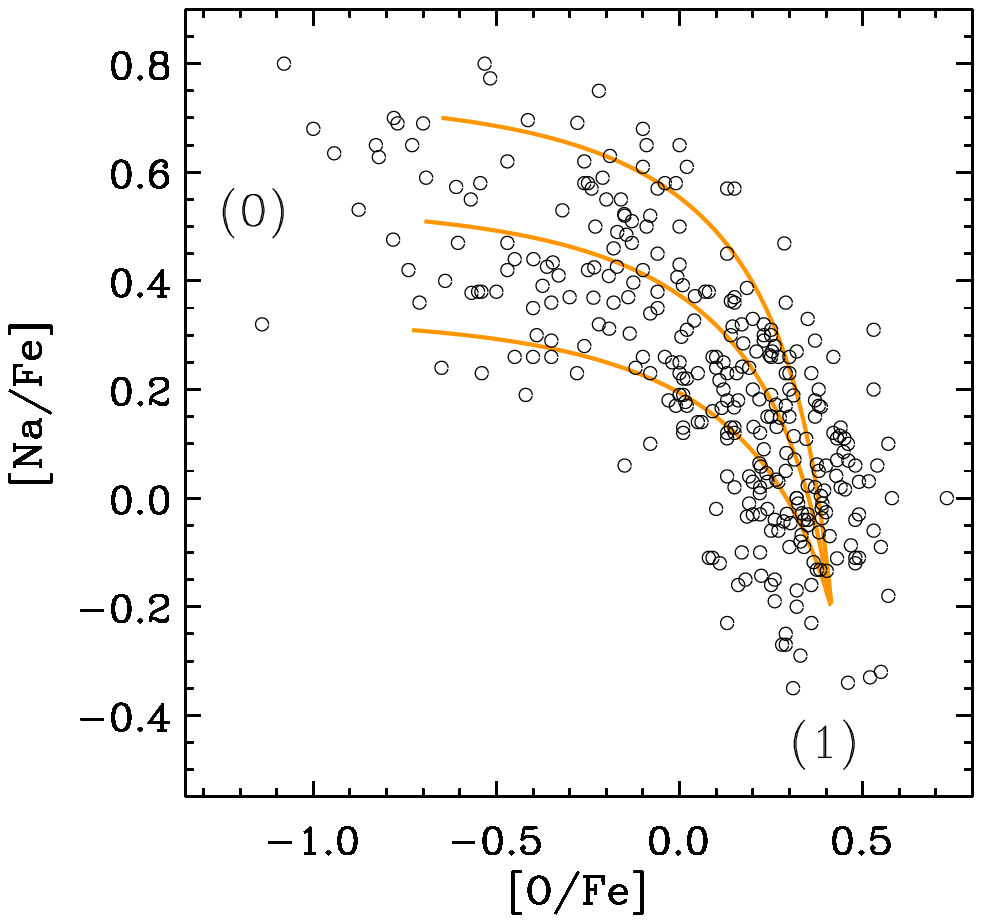,width=0.40\textwidth}
\psfig{figure=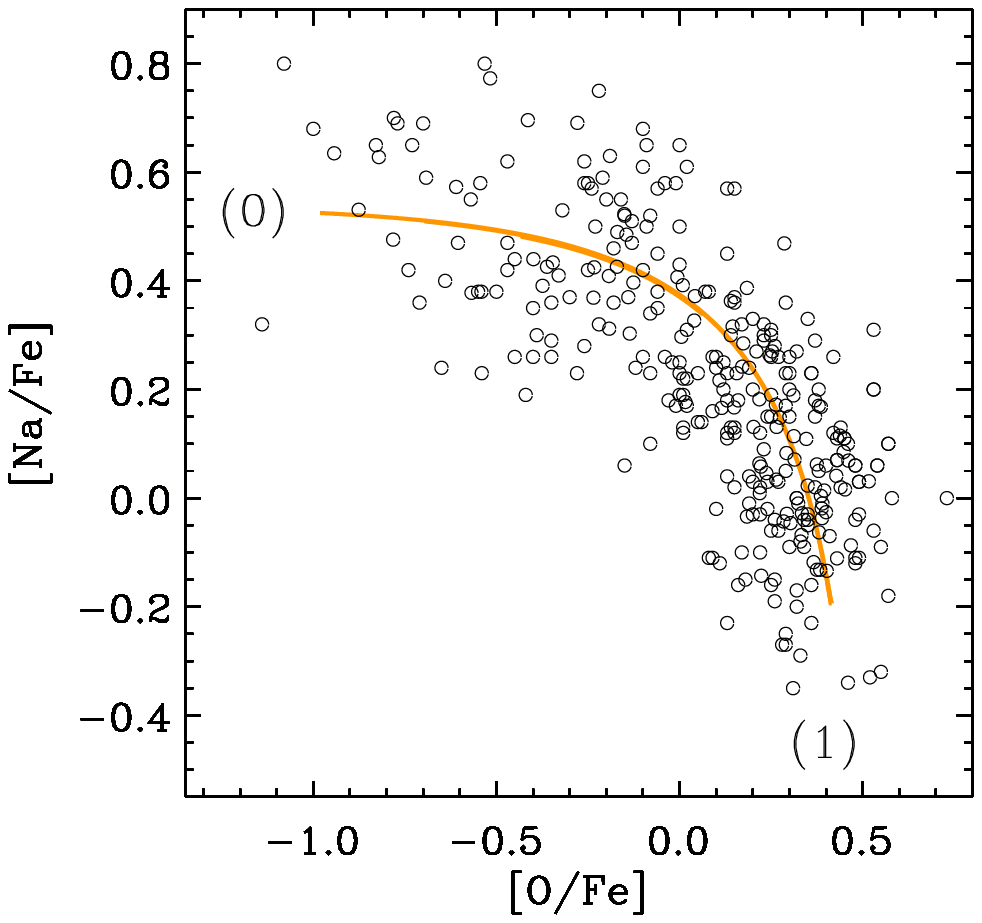,width=0.40\textwidth}
\psfig{figure=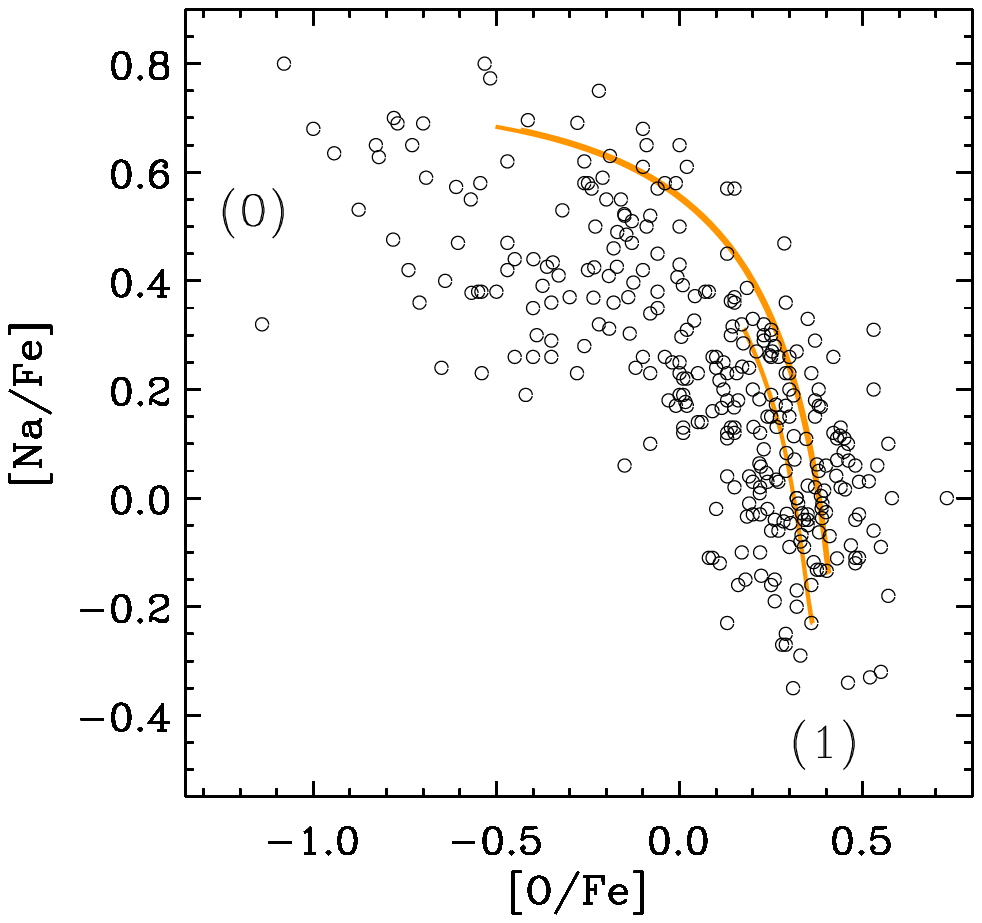,width=0.40\textwidth}
\end{center}    
\caption{Representative model predictions for the O-Na anti-correlation. 
The first generation of stars are born with peculiar chemical properties 
(i.e., low [O/Fe] and high [Na/Fe]). Once the first SNeII start to 
explode, polluting and expanding the inner region, the chemical 
properties of the forming stars evolve toward ``normal'' values typical 
of pure SNeII pollution (i.e., high [O/Fe] and low [Na/Fe]). Upper left panel:
the three 
lines represent models with different initial conditions - i.e., models 
in which the number of intermediate-mass AGB stars and the initial 
radius of the inner pre-polluted region has been changed to match the 
observational scatter observed in different GCs (small circles, see text 
for more details about the data sets). The solid lines represent models 
with the combinations of values of N$_{AGB}$=100, 160, 250 and R$_{\rm in}$=40, 
30 and 20 pc with the former (latter) values representing the 
lower (upper) line.
Upper right panel: 
The three lines correspond to the models with the same inner radius 
($R_{\rm in}$=30 pc), but with different number of AGB stars: N$_{AGB}$=100, 160, and 250, from lower to
upper line, respectively.
 Bottom left  panel: 
 The three lines reflect the effect of changing the inner radius, keeping the number of AGB stars fixed to 
  N$_{AGB}$=160.
Bottom right panel: Effect of changing the [Fe/H]$_{\rm ISM}$ keeping the 
rest of the parameters fixed. From top  to bottom 
[Fe/H]$_{\rm ISM}$=$-3.30$, $-2.55$ and $-1.40$.}
\label{fig:nao}  
\end{figure*} 
 
In Table~\ref{tab:yields}, we summarise the mean yields for 
intermediate-mass AGB stars (see the complementary Table~3 of Paper~I, 
for the elements studied in our earlier work) calculated by averaging 
over a Salpeter (1955) IMF, in the mass range 4$-$7~M$_{\odot}$, the 
Karakas \& Lattanzio (2007)\nocite{karakas2007} yields with Z=0.0001 (we refer to this 
model as the ``reference model''). There are several Fe-peak elements 
which were not considered by these authors (i.e., Ca, Ti, V, Cu) and, 
for this work, we have therefore assumed that these elements are not 
synthesised in significant quantities in intermediate-mass AGB 
stars.\footnote{We note here in passing that our preliminary work now 
suggests that low-metallicity, massive, AGB stars can produce a 
substantial amount of copper, ie.,  one 5~M$_\odot$, Z=0.004 model we have 
generated produced [Cu/Fe]$\sim$$+$0.8.  A larger grid of models will be 
required before we can assess its global importance within the context 
of our GC modeling efforts.}

For SNe~Ia we use the yields of \citet{iwamoto1999} and explore the 
impact of various stellar physics treatments, including slow 
deflagration (W7) and delayed detonation models (WDD1 and CDD1). While 
SNe~Ia produce small amounts of light elements (relative to SNe~II), 
they can produce significant amounts of heavier $\alpha$-elements (e.g., 
Si, Ca, and Ti) and Fe-peak elements. The specific yield of the latter 
amount depends upon the deflagration speed and ignition densities, with 
the different models producing yields that span a factor of four (see 
Table~\ref{tab:yields} and, more importantly, the original paper of 
Iwamoto et~al. 1999 for details).
 
\begin{table*}  
\centering
\begin{minipage}{165mm}  
\caption{Mean SNe~II, SNe~Ia, and intermediate-mass AGB yields in solar 
masses. The SNe~II entry corresponds to the adopted ``average'' Model 
listed at the end of Table~1.  The SNe~Ia yields corresponds to the 
models described by \citet{iwamoto1999}. The mean AGB yields (from 
Karakas \& Lattanzio 2007) are averaged over the progenitor mass range 
4$-$7~M$_{\odot}$ assuming a Salpeter (1955) IMF.}
\label{tab:yields}    
\begin{tabular} {|l|c|c|c|c|c|c|c|c|c|c|c|c|c|c|c|}  
\hline \\   
SNe type &$\;\;\;\;$ Fe $\;\;\;\;$&$\;\;\;\;$ F $\;\;\;\;$&$\;\;\;\;$ Si $\;\;\;\;$&  
$\;\;\;\;$ Ca $\;\;\;\;$&
$\;\;\;\;$ Ti $\;\;\;\;$&$\;\;\;\;$ V $\;\;\;\;$&$\;\;\;\;$ Co $\;\;\;\;$&$\;\;\;\;$ Ni  
$\;\;\;\;$
& $\;\;\;\;$ Cu $\;\;\;\;$ \\  
\hline 
SNe II (Model)   & 9.00e-2  & 3.00e-5   & 1.28e-1 & 9.40e-3 & 2.23e-4 & 1.33e-5 & 2.45e-4  
& 5.33e-3 & 1.33e-5 \\ 
\hline 
SNIa (W7)        & 7.49e-1  & 5.67e-10 & 1.56e-1 & 1.19e-2& 3.43e-4  & 7.49e-5 & 1.04e-3  
& 1.26e-1 & 3.00e-6  \\ 
SNIa (WDD1)      & 6.72e-1  & 1.70e-9  & 2.74e-1 & 3.10e-2& 1.13e-3  & 1.33e-4 & 3.95e-4  
& 3.40e-2 & 6.92e-6  \\ 
SNIa (CDD1)      & 6.48e-1  & 5.83e-10 & 2.79e-1 & 3.18e-2& 8.18e-4  & 1.11e-4 & 2.91e-4  
& 3.50e-2 & 7.28e-7  \\ 
\hline 
AGB (model) & 2.27e-5  & 4.98e-9  & 4.87e-5 &    0.  &   0.     &   0.    &   7.93e-8     
& 3.07e-6   & 0.   \\ 
\hline  
\end{tabular}    
\end{minipage} 
\end{table*}

\section{Model} 
\label{sec:model} 

The details of the models are described in Paper~I; here, we briefly 
review the main characteristics.  The model has two phases: first, the 
localised effect of intermediate-mass AGB field stars and a single SNIa 
explosion produce an inhomogeneous pollution that is added to an 
interstellar medium that was previously enriched by very low metallicity 
SNe~II. Since the peak of the SNIa rate has a timescale 
\citep[$\sim$70$-$80~Myr; e.g.][]{matteucci2001, mannucci2006} 
comparable with the lifetime of a 5~M$_{\odot}$ star 
\citep{schaller1992, karakas2007} such pre-enrichment is not \it a 
priori \rm unreasonable (Paper~I provides a detailed discussion of the 
initial conditions). The formation of a proto-GC takes place inside this 
chemically-peculiar region; the stars formed during this phase will have 
``peculiar'' chemical properties compared with ``normal'' field stars of 
the same metallicity.

As star formation proceeds, new SNe~II explode, polluting the gas with 
the product of their nucleosynthesis. This represents the second phase 
of the model. The stars formed in this phase will have chemical 
properties that are more similar to the field stars of the same 
metallicity (henceforth referred to as ``normal''). Eventually, the 
occurrence of SNe~II explosions quenches further star formation and the 
GC evolves passively.
  
\begin{figure*} 
\begin{center} 
\psfig{figure=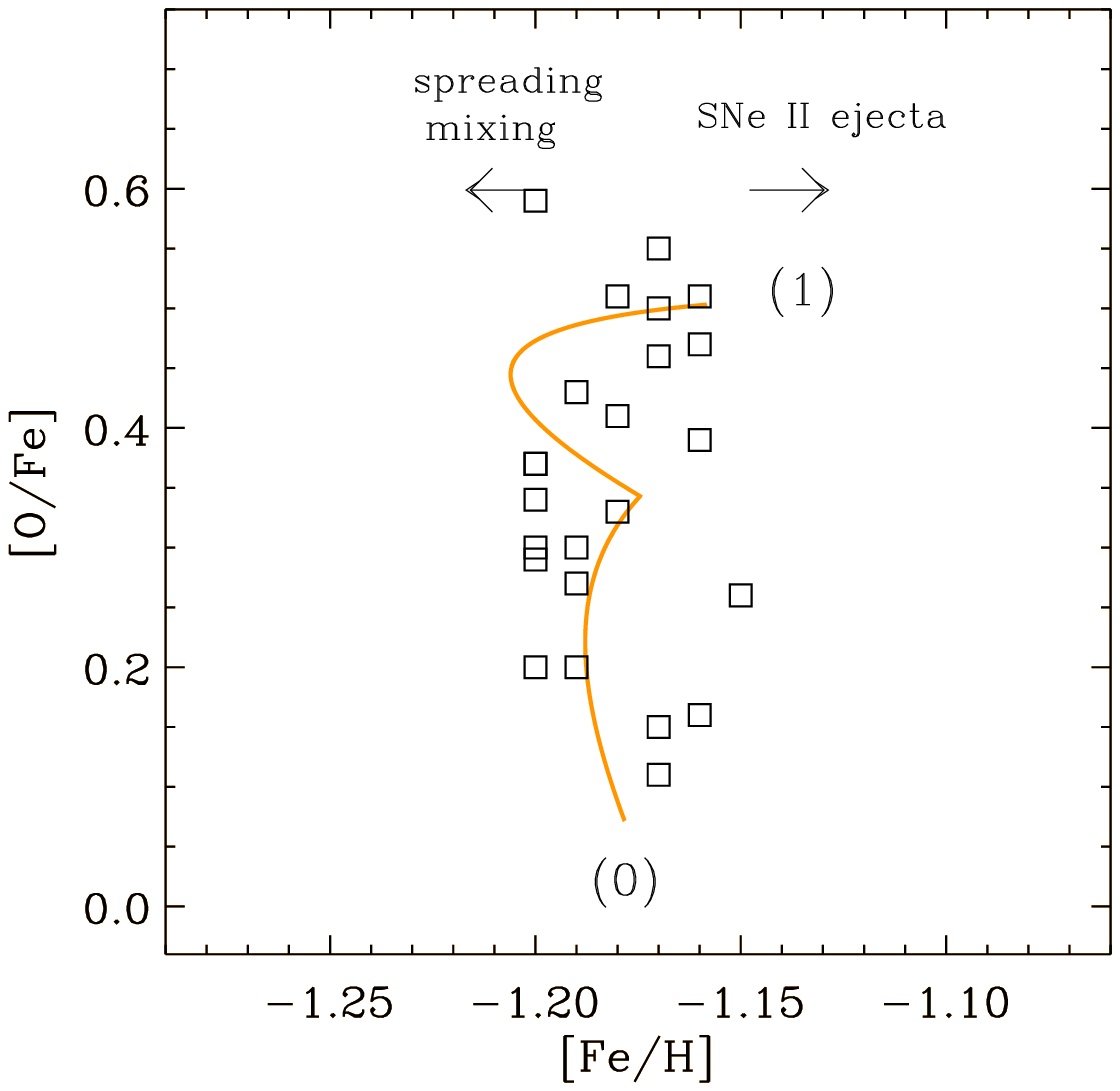,width=0.40\textwidth} 
\psfig{figure=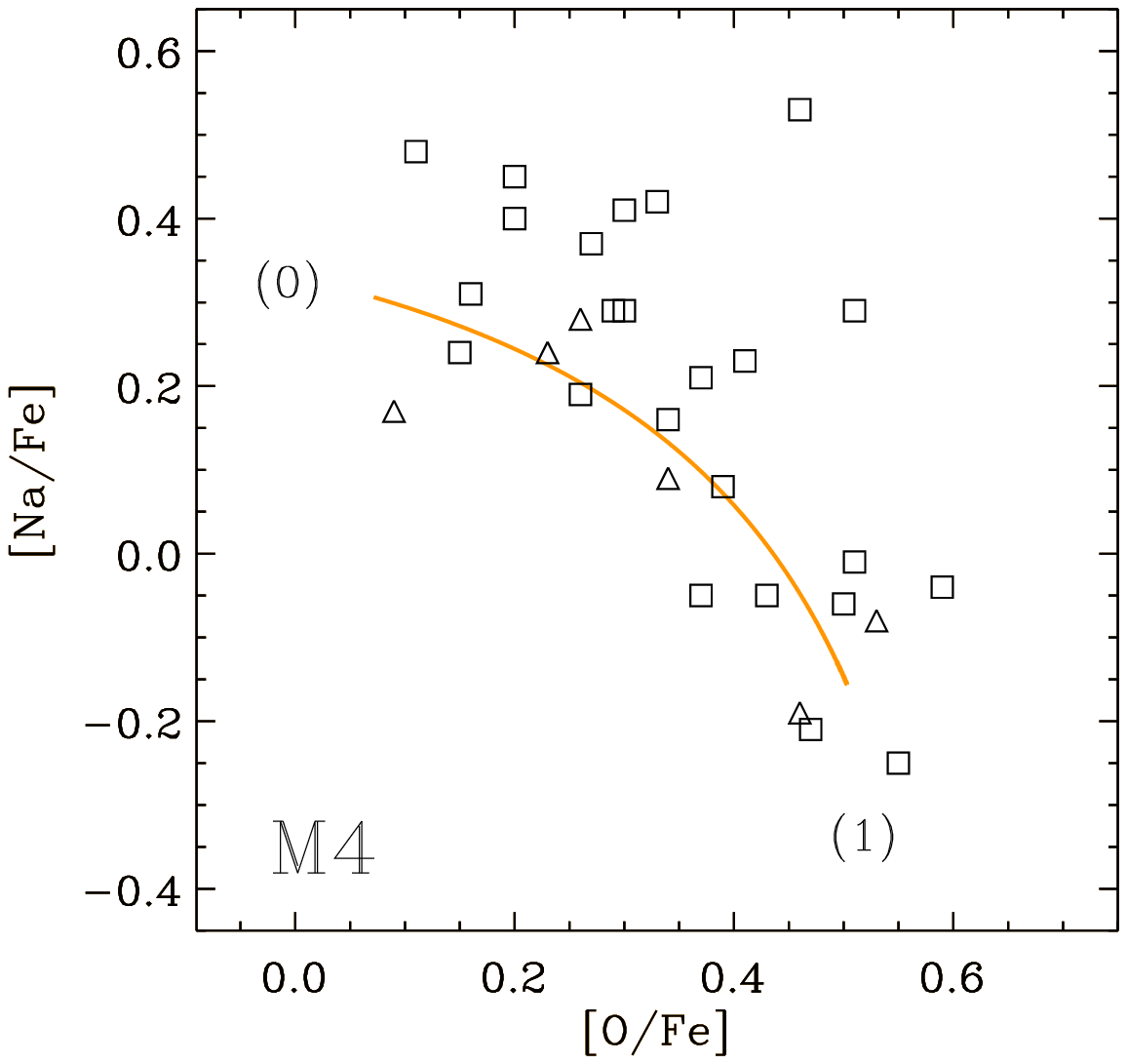,width=0.40\textwidth} 
\psfig{figure=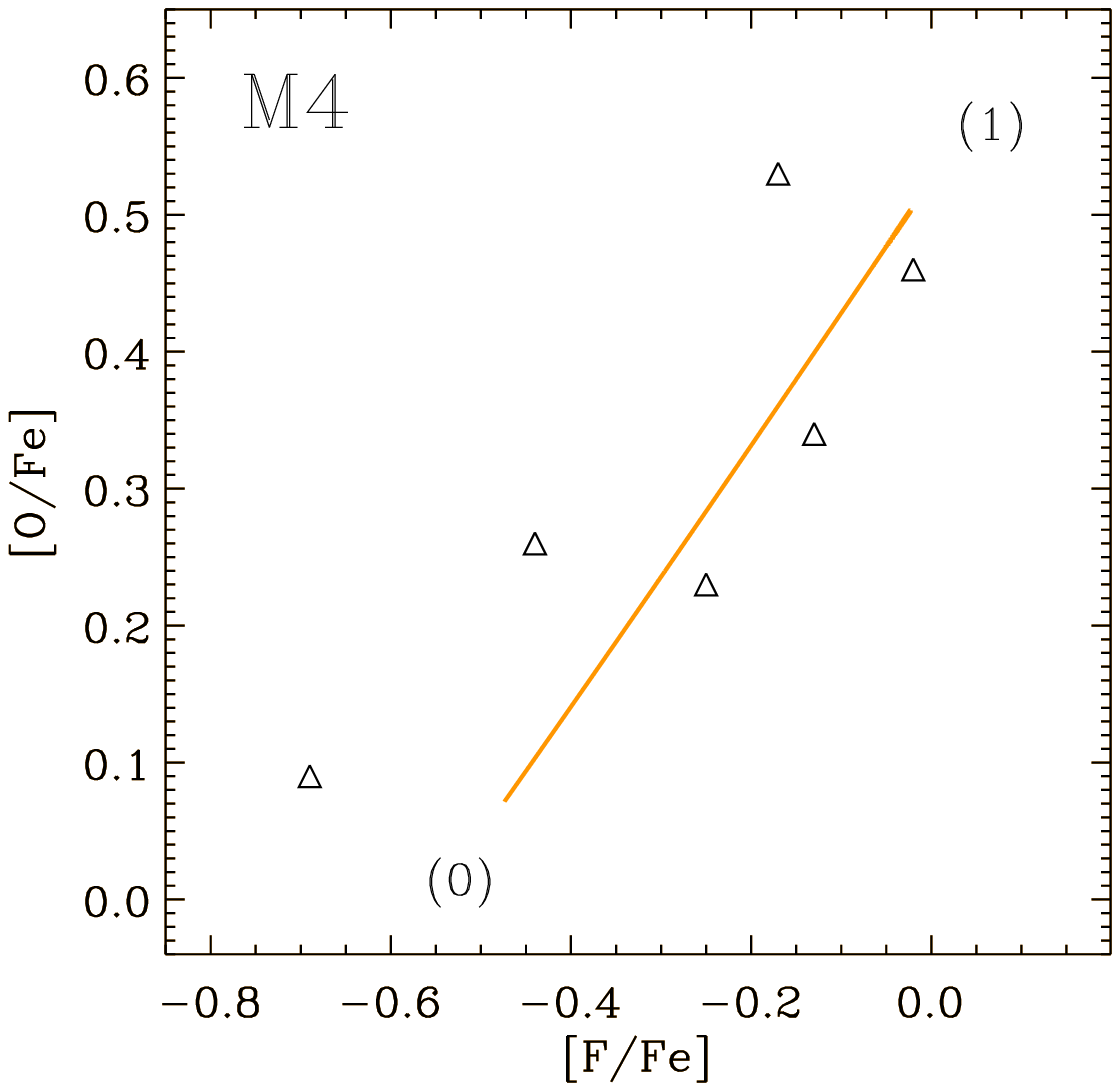,width=0.40\textwidth} 
\psfig{figure=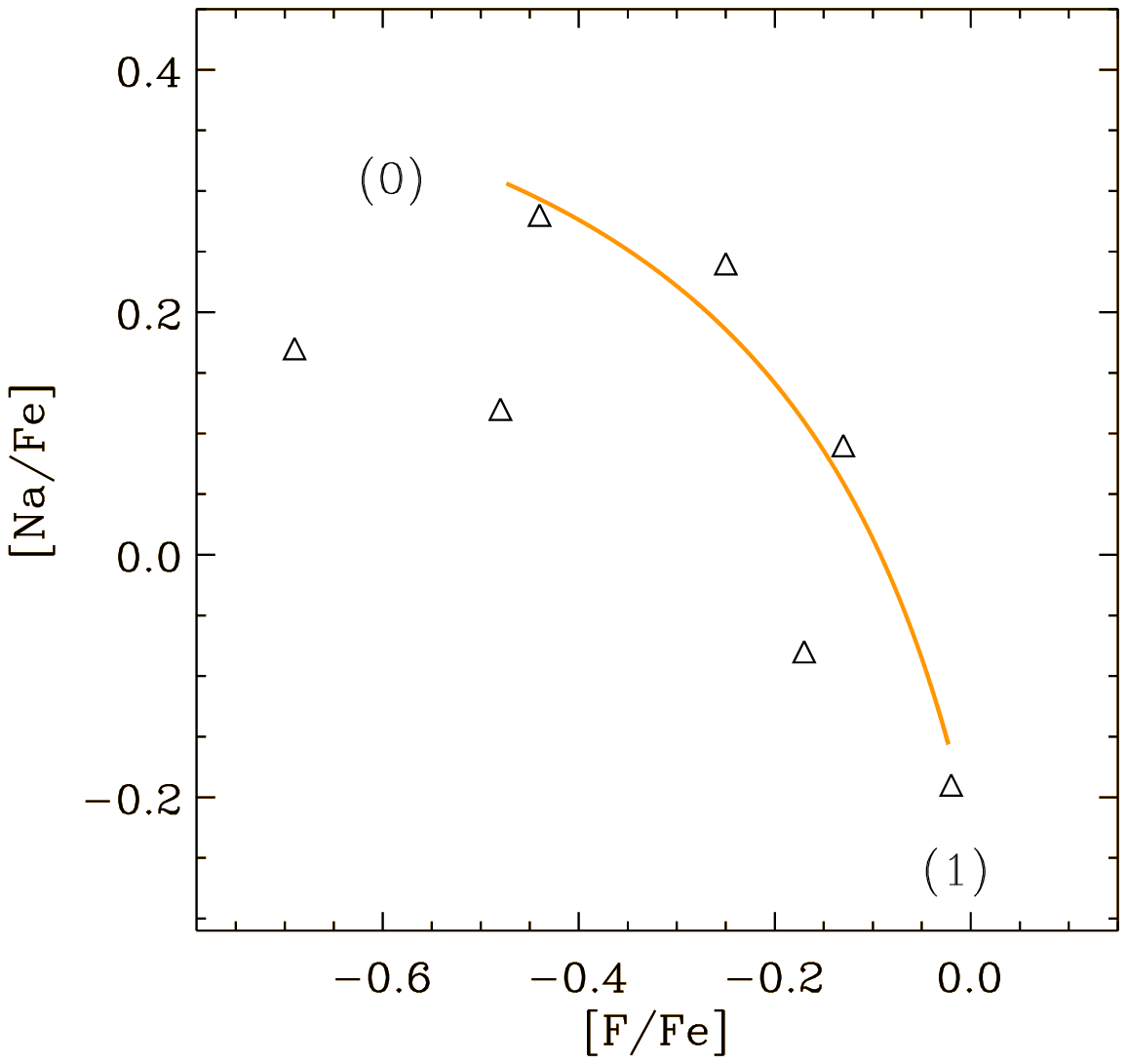,width=0.40\textwidth} 
\end{center} 
\caption{Evolution of various element ratios for the model for globular 
cluster M4; comparison observational data from 
\citet[][squares]{ivans1999} and \citet[][triangles]{smith2005}. 
Upper-left panel: evolution of [O/Fe] versus [Fe/H]. Upper-right panel: 
evolution of [Na/Fe] versus [O/Fe]. Lower-left panel: evolution of 
[O/Fe] versus [F/Fe]. Lower-right panel: evolution of [Na/Fe] versus 
[F/Fe].}
\label{fig:M4} 
\end{figure*} 

During the evolution there are two mechanisms that govern the chemical 
properties of the forming stars. We showed in Paper~1 that the two 
mechanisms act together to regulate the [Fe/H] abundance, which is kept 
essentially constant throughout a cluster's evolution. Furthermore, the 
general chemical properties of the stars forming in this region evolve 
from ``peculiar'' to that more typical of essentially ``pure'' SNe~II 
enrichment, which is the opposite to what is normally assumed in 
self-enrichment models.

The final chemical properties of the GC are controlled by three 
parameters: the size of the inner region where the SNIa is confined 
(R$_{\rm in}$), the initial number of AGB stars ($N_{\rm AGB}$), and the 
abundance of the interstellar medium (ISM) of the halo ([Fe/H]$_{\rm 
ISM}$). Table~\ref{tab:gc_mod} lists these parameters for the globular 
clusters examined in this paper.  We can see that for these models, 
there is an inverse correlation between the value of the surrounding 
[Fe/H]$_{\rm ISM}$ and the confinement radius of the SN~Ia, R$_{\rm 
in}$.\footnote{Admittedly, the correlation is not a particularly strong 
one, particularly when the two models from Paper~I (bottom two entries 
of Table~3) are included.} Such an inverse correlation is consistent 
physically with the expected inverse correaltion between metallicity and 
cooling efficiency. The higher the metallicity of the confining ISM, the 
more efficient the cooling and energy dissipation, and the more confined 
the SN remnant's expansion will be. A factor of 100 increase in 
metallicity would make the radius at which the remnant merges with the 
ISM a factor of three smaller and, therefore, the enrichment will be 
spread over a smaller volume \citep{gibson1994}.
 

To illustrate the effect of the model free parameters on the observed O-Na 
anti-correlation we show, in Fig.~\ref{fig:nao} models changing 
the number of AGB polluters, the initial radius (R$_{\rm in}$ and the initial 
iron content of the interestellar medium [Fe/H]$_{ISM}$, together
with a collection of observational data for different GCs \citep{kraft1993, 
sneden1997, ivans2001, ramirez2003, sneden2004, cohen2005, yong2005, carretta2006, sneden2004, koch2008}. 
We also show three different models with (with $N_{\rm AGB}$=100, 160, 250, and R$_{\rm in}$=40, 30, and 
20~pc. It 
can be seen that appropriate choices for the initial conditions are able
to reproduce both the trend and the associated ``scatter'' in the 
observed O-Na anti-correlation.
 
\section{Very metal-poor and metal-rich GCs} 
 
\subsection{The case of very metal-poor GCs: M~15 and NGC~6397} 
 
\begin{figure*} 
\begin{center} 
\psfig{figure=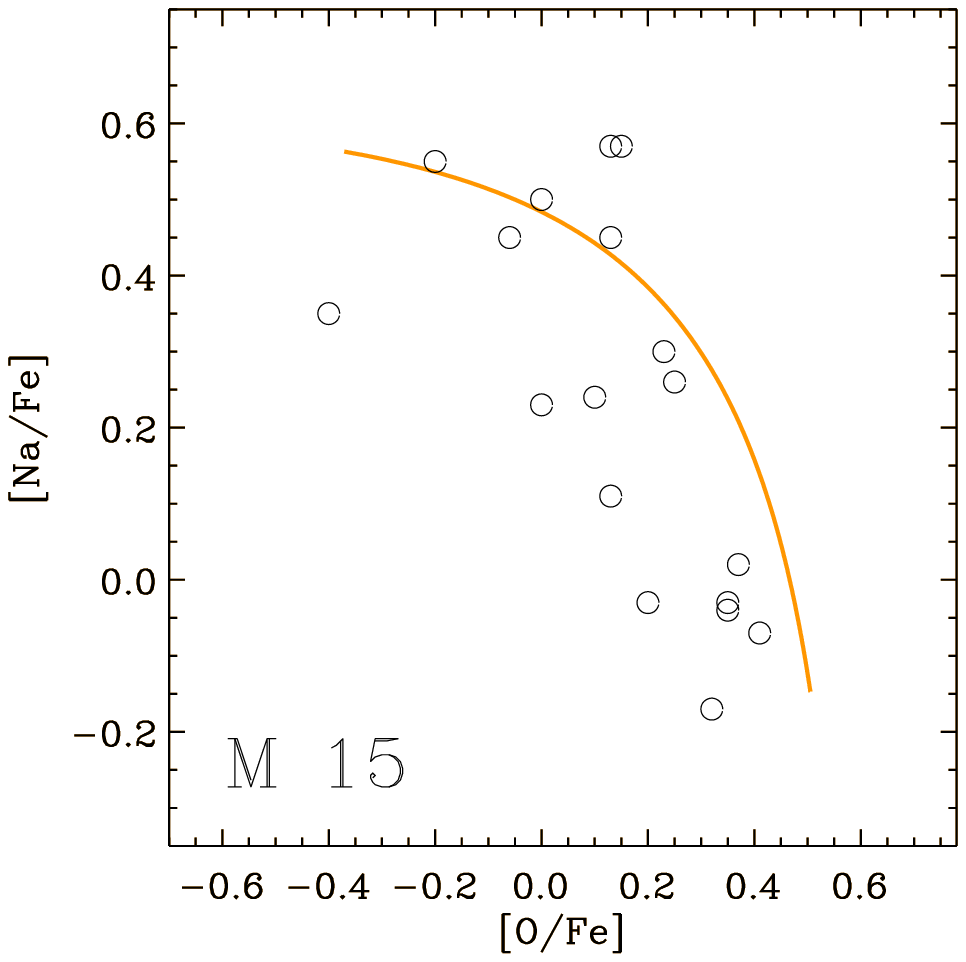,width=0.40\textwidth} 
\psfig{figure=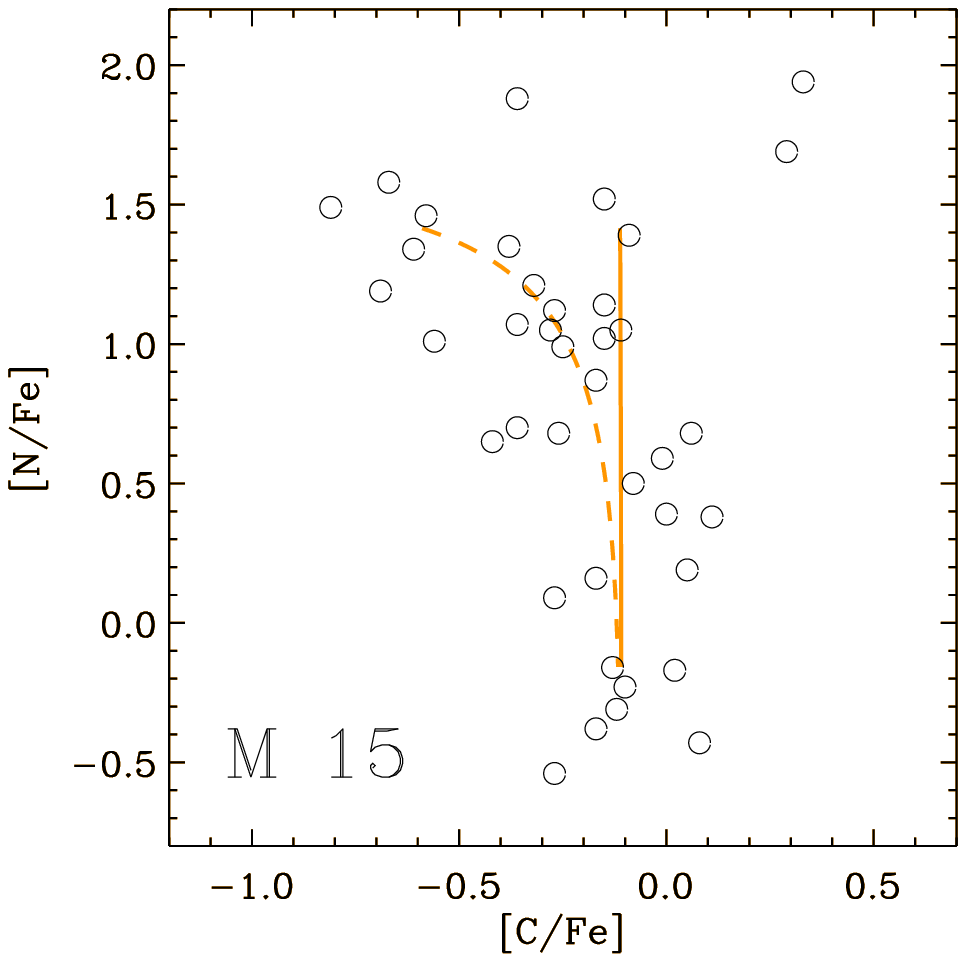,width=0.40\textwidth} 
\psfig{figure=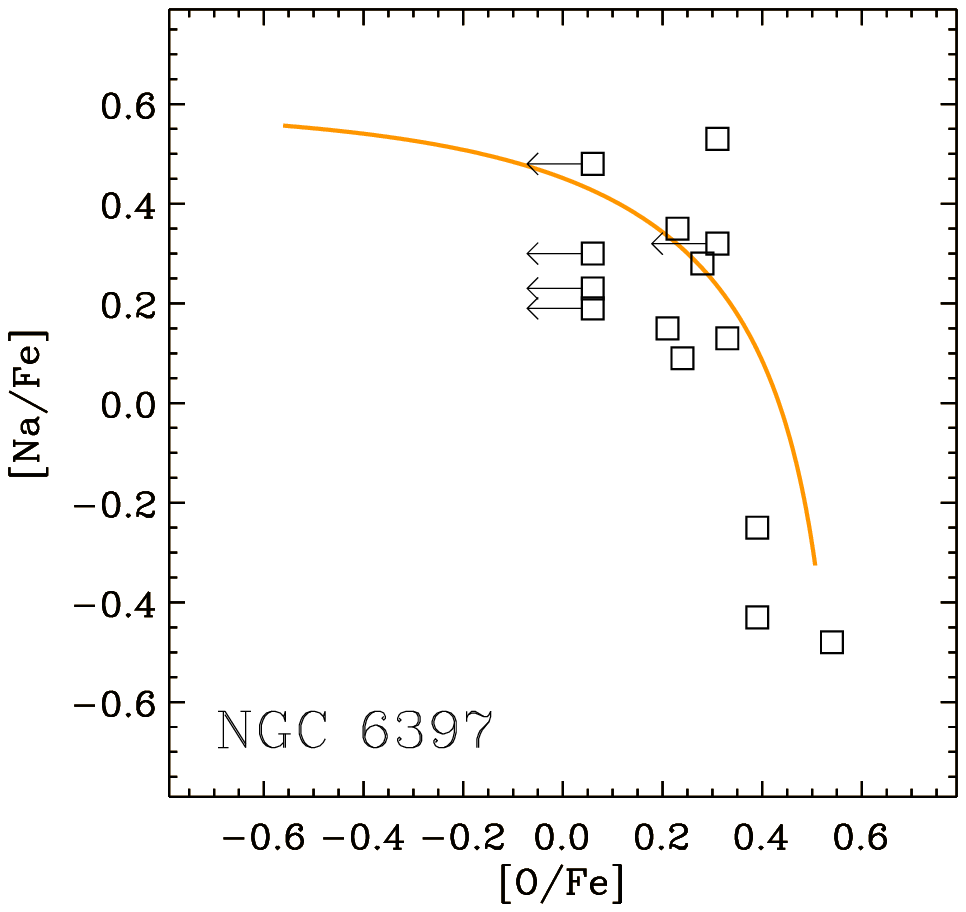,width=0.40\textwidth} 
\psfig{figure=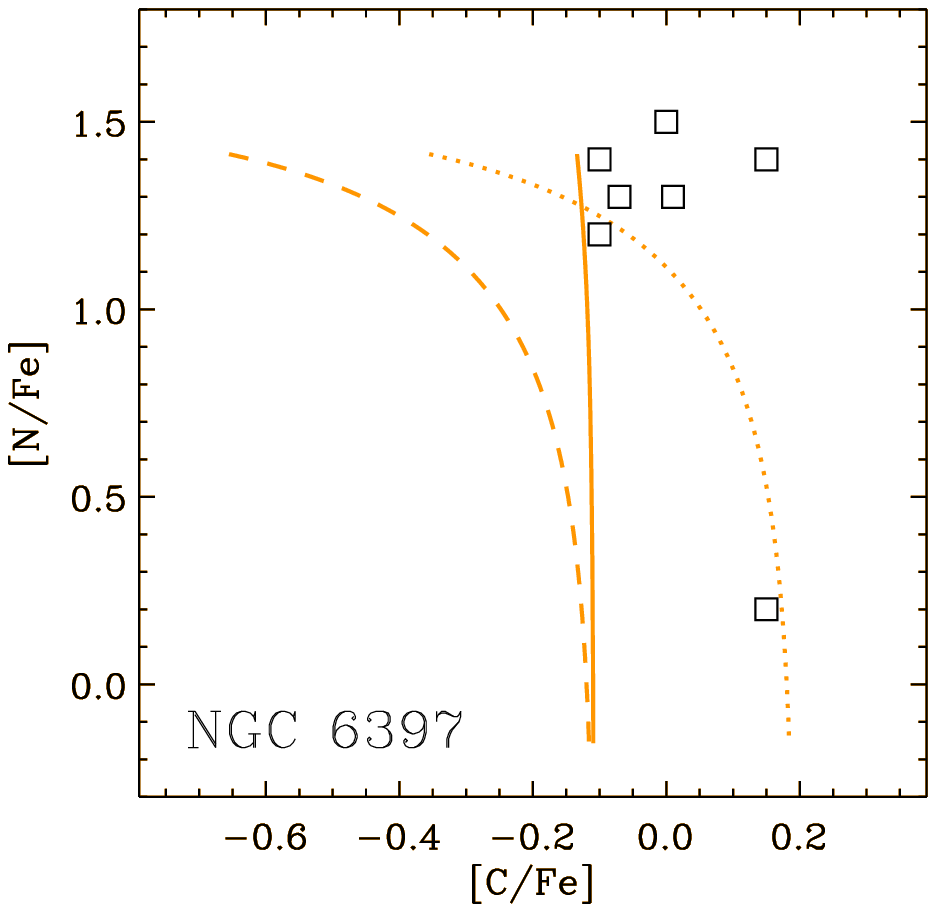,width=0.40\textwidth} 
\end{center}    
\caption{Upper-left panel: evolution of [Na/Fe] versus [O/Fe] for 
the model of  M15 (solid line) plotted against the 
data by \citet{sneden1997}. Upper-right 
panel: evolution of [N/Fe] versus [C/Fe] for the same model 
(reference model, solid line) plotted against the data by
\citet{cohen2005}. Dashed line indicate the effect of 
reducing the production of C by AGB stars by a factor of four.
Lower panel: Same, but showing the 
evolution of the NGC~6397 model, plotted together with the
 observational data by \citet{carretta2005}. Note that in this case most 
of the lowest [O/Fe] abundances are upper limits. The dotted line in the 
lower-right panel corresponds to the same evolution of the dashed line 
(reference model with a reduced production of C by AGB stars) but offset by 0.3~dex
(see text for details).}
\label{fig:M15} 
\end{figure*}

In Figure~\ref{fig:M15} we show the [Na/Fe]--[O/Fe] and the 
[N/Fe]--[C/Fe] evolution predicted by our models for M15 and NGC~6397 
together with the corresponding observational datasets of 
\citet{sneden1997}, \citet{cohen2005}, and \citet{carretta2005}. 
Table~\ref{tab:gc_mod} lists the model parameters inferred which 
best-fit the overall [Fe/H] content of NGC~6397 and M15. To fit the 
abundances for these clusters, a larger value of R$_{\rm in}$ and lower 
value of [Fe/H]$_{\rm ISM}$ were needed, relative to those employed for 
the more metal-rich GCs. The models recover the O-Na anti-correlation 
for both clusters, but are less successful in doing the same for the C-N 
anti-correlation. This was discussed in Paper~I, where we showed that 
while the the main contribution to the production of carbon and nitrogen 
comes from intermediate-mass AGB stars, the amount of nitrogen produced 
by these stars is much larger than that of carbon. The result is that 
while the initial AGB production of carbon is offset by the Fe deposited 
by the SN~Ia, the [N/Fe] ratio can reach values as high as 1.5~dex. For 
this reason, while the carbon content remains practically constant 
during the evolution (solid lines), the nitrogen content varies by more 
than an order of magnitude (see right panels of Figure~\ref{fig:M15}). As 
a consequence, our reference models (solid lines) are able to reproduce 
the observed spread in N while keeping C constant. However, in order to 
obtain an anti-correlation between the relative abundance of these 
elements, the production of C by AGB stars needs to be reduced by a 
factor of four (dashed lines of  Figure~\ref{fig:M15}).

It can also be seen that the absolute values of [C/Fe] for NGC~6397 are 
somewhat difficult to reproduce in our model, particularly at low [N/Fe] 
values. However, the comparison with carbon and nitrogen measurements 
need to be done with caution. The abundances in NGC~6397 have been 
derived from spectra of slightly evolved stars (subgiants). The surface 
abundances of these elements are affected by internal mixing and may not 
reflect the compositon of the ISM when those stars formed 
\citep[see][and reference therein]{smith1992, 
charbonnel1994,charbonnel1995, denissenkov2000, weiss2000, gratton2004}. 
The carbon abundance measurements of M15 and NGC~6397 differ by 
$\sim$0.3~dex (see right panel of Figure~\ref{fig:M15}). This is very 
difficult to understand given the similarities in their overall 
metallicity and nitrogen content. We have assumed here that this offset 
is artificial and perhaps reflects a systematic error in the carbon 
measurements of NGC~6397, but admit that this remains, for now, just an 
assumed interpretation. If we apply an offset to the carbon abundances 
of the NGC~6397 stars to match the abundances of M15 stars, our model 
agrees better with the observations (Fig.~\ref{fig:M15}).

\begin{figure*}     
\begin{center}     
\psfig{figure=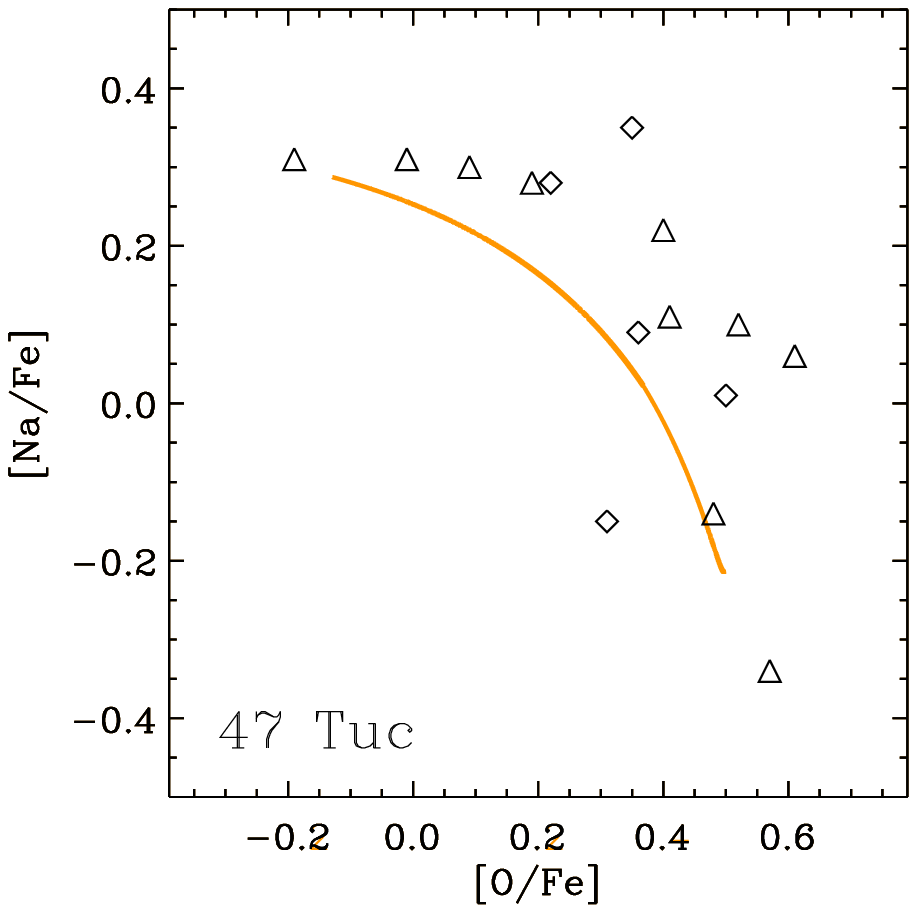,width=0.40\textwidth}
\psfig{figure=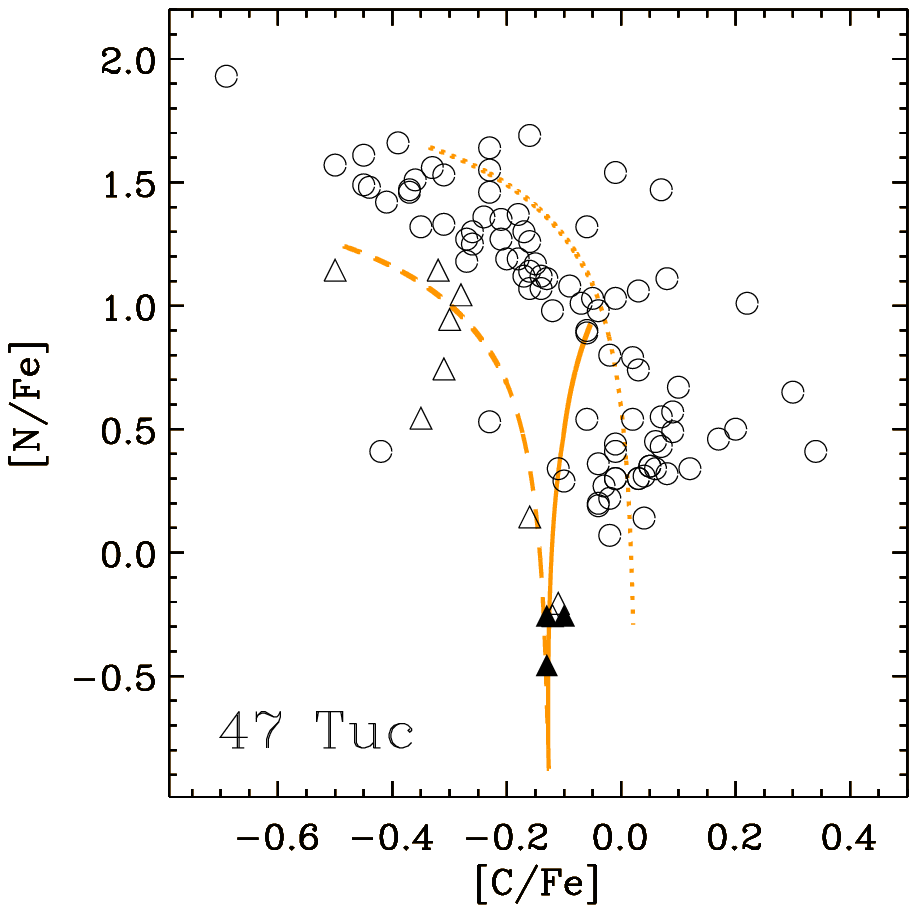,width=0.40\textwidth}
\end{center}    
\caption{Left panel: evolution of [Na/Fe] versus [O/Fe] for the reference model 
for 47~Tuc plotted against the observational data collected by 
\citet[][triangles]{carretta2005} and \citet[][diamonds]{alves2005}. 
Right panel: evolution of [N/Fe] versus [C/Fe] for the same model 
plotted against the values presented in 
\citet[][triangles]{carretta2005}, \citet[][circles]{briley2004a}. Dashed line
indicate a model where the production of C by AGB stars has been reduced in 
a factor of 4 with respect to the reference model. In 
the case of \citet{carretta2005} open triangles refer to sub-giant stars 
while filled triangles refer to dwarf stars. The dotted line in the 
right panel corresponds to the dashed-line model offset by 0.15~dex in 
carbon and 0.40~dex in nitrogen.}
\label{fig:47Tuc} 
\end{figure*}

\begin{table}  
\centering  
\begin{minipage}{80mm}  
\caption{Initial conditions of the models for the different globular clusters} 
\label{tab:gc_mod} 
\begin{tabular} {|l|c|c|c|c|c|}  
\hline  
   & [Fe/H]$_{\rm ISM}$ & [Fe/H]$_{\rm in}$ & R$_{\rm in}$ (pc) & $N_{\rm AGB}$ \\  
\hline 
M~15        &  $-3.30$     &  $-2.35$      &     65         &     150   \\ 
NGC~6397    &  $-3.20$     &  $-2.05$      &     49         &     150   \\ 
NGC~6752    &  $-2.55$     &  $-1.60$      &     36         &     250   \\ 
M~4         &  $-1.65$     &  $-1.15$      &     29         &     130   \\ 
47~Tuc      &  $-1.40$     &  $-0.75$      &     20         &     140   \\ 
\hline 
M~13        &  $-3.50$     &  $-1.50$      &     31         &     170   \\ 
NGC~2808    &  $-2.85$     &  $-1.10$      &     24         &     180   \\ 
\hline 
\end{tabular}    
\end{minipage}    
\end{table}

\begin{figure*}     
\begin{center}     
\psfig{figure=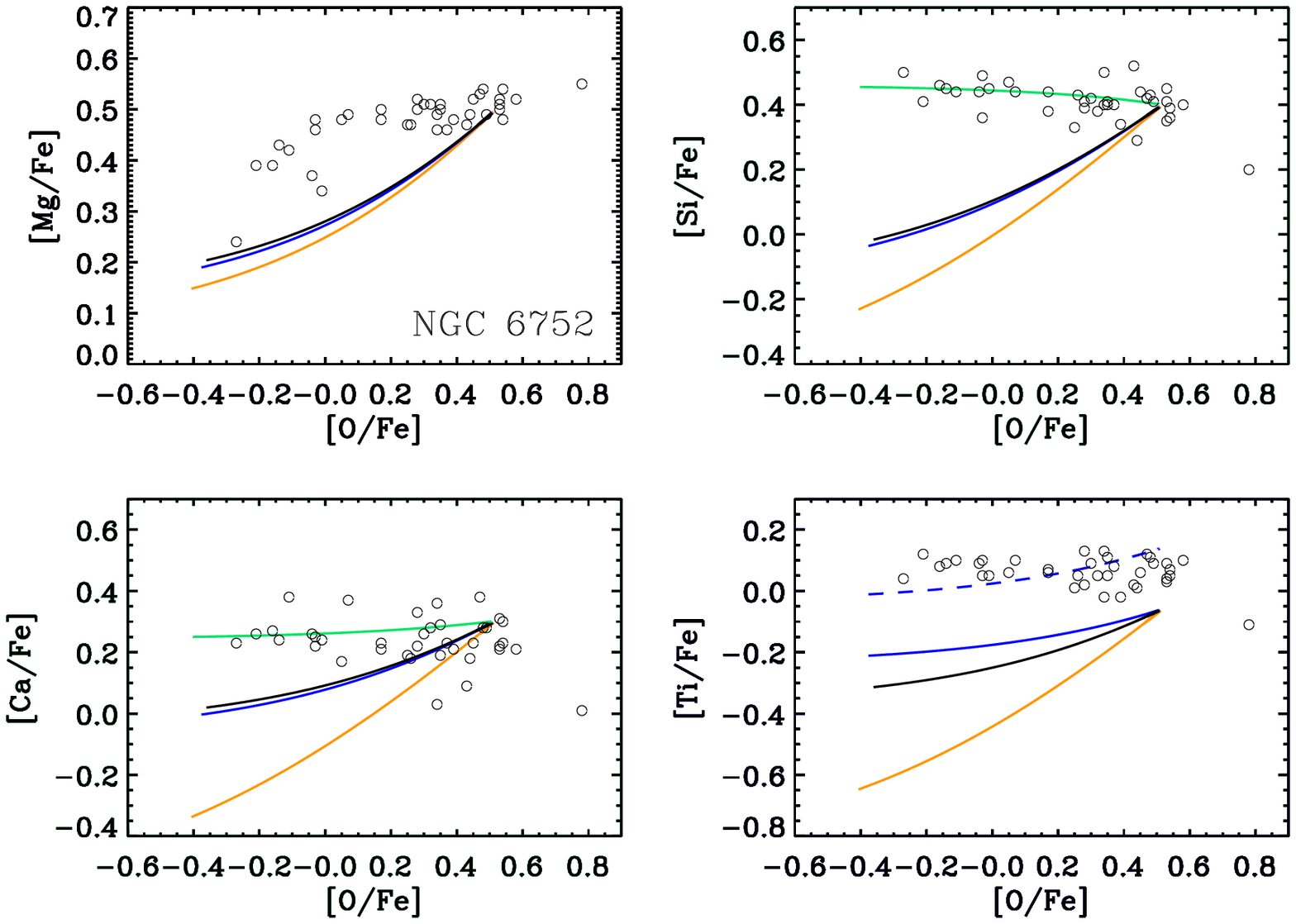,width=0.90\textwidth} 
\end{center}    
\caption{Evolution of different $\alpha$-elements ([Mg/Fe], [Si/Fe], 
[Ca/Fe] and [Ti/Fe]) versus [O/Fe] for the case of NGC~6752. Open 
circles correspond to the observational dataset from \citet{yong2005}.  
The colour-coded lines correspond to models with different SNe~Ia 
yields.  Orange lines: W7; blue lines: WDD1; black lines: CDD1. The 
green lines represent a model in which the Si and Ca yields have been 
increased of a factor of three and two, respectively, compared with the 
yields of model WDD1. The dashed blue line in the lower-right panel is 
the corresponding blue solid line, offset by 0.2 dex.}
\label{fig:NGC6752_alpha}  
\end{figure*}

\subsection{The case of metal-rich GCs: 47~Tuc} 
 
As a final test for the evolution of the abundances of light elements we 
consider the case of the metal-rich globular cluster 47~Tuc 
([Fe/H]$=-0.67$) \citep[e.g.,][]{carretta2004}. In order to match its 
chemical properties we assume that the initial SN~Ia was quite localised 
(R$_{\rm in}$=20 pc) and that the Fe content of the ISM of the halo at 
the epoch of formation was higher than for the metal-poor GC 
([Fe/H]$_{\rm ISM}=-1.40$). This value is in the range of iron 
metallicities encountered in the Galactic halo ($-4.0\ge$[Fe/H]$\ge$+0.0 
peaking near [Fe/H]$\simeq -1.7$ \citep[e.g.,][]{ryan1991}).
 For the specific case of 47~Tuc we have used the $Z$=0.004 
yields, consistent with its higher initial metal content.

The evolution of the O-Na and C-N anti-correlations are plotted in 
Fig.~\ref{fig:47Tuc} together with the observational data of 
\citet[][]{carretta2005}, \citet[][]{alves2005}, and 
\citet[][]{briley2004a}. Again the anti-correlations are better 
reproduced once we assume that the carbon production in AGB stars is 
reduced by a factor of four compared with the theoretical values of 
\citet{karakas2007}. There is also an evident offset of 0.15 and 0.4 dex 
between the measurements of carbon and nitrogen derived by 
\citet{carretta2005} and \citet{briley2004a}. While it is beyond the 
scope of this paper to understand the origin of these offsets, our 
assumed yields are in agreement with the values obtained by 
\citet{carretta2005} but, again, the model has no problem in being able 
to reproduce the trend from \citet[][dotted line]{briley2004a}.

\subsection{[Fe/H] evolution}

\begin{figure*} 
\begin{center} 
\psfig{figure=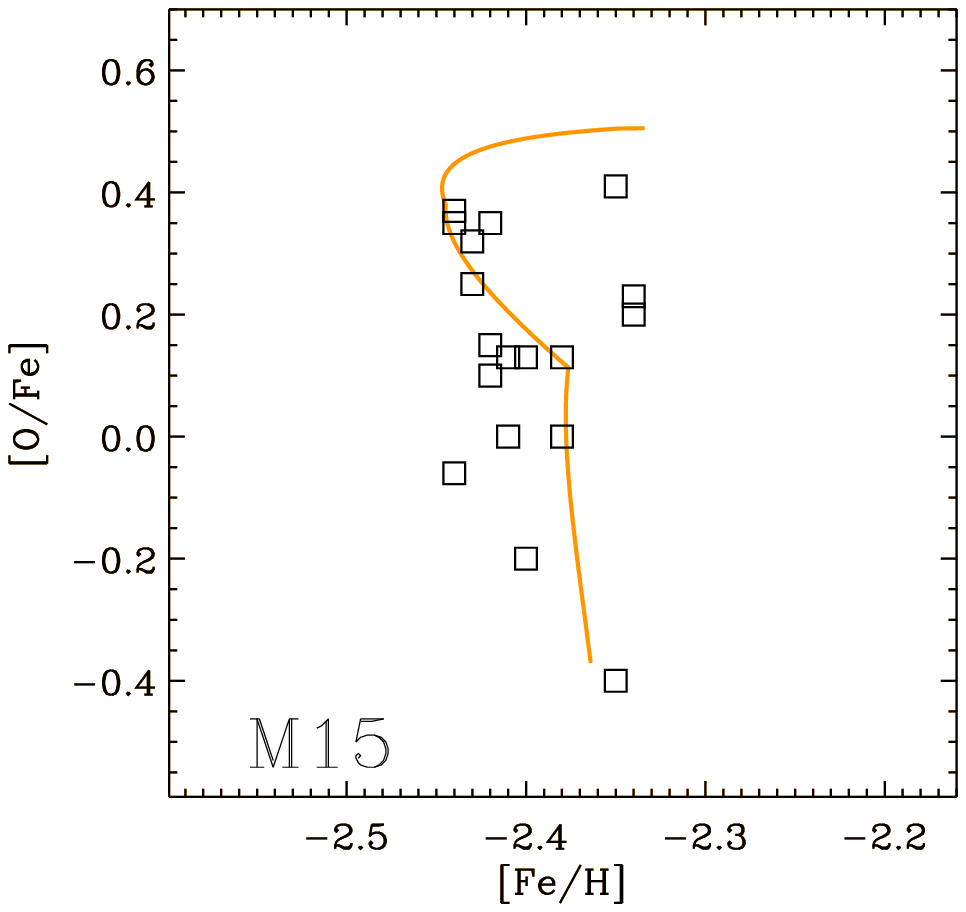,width=0.40\textwidth} 
\psfig{figure=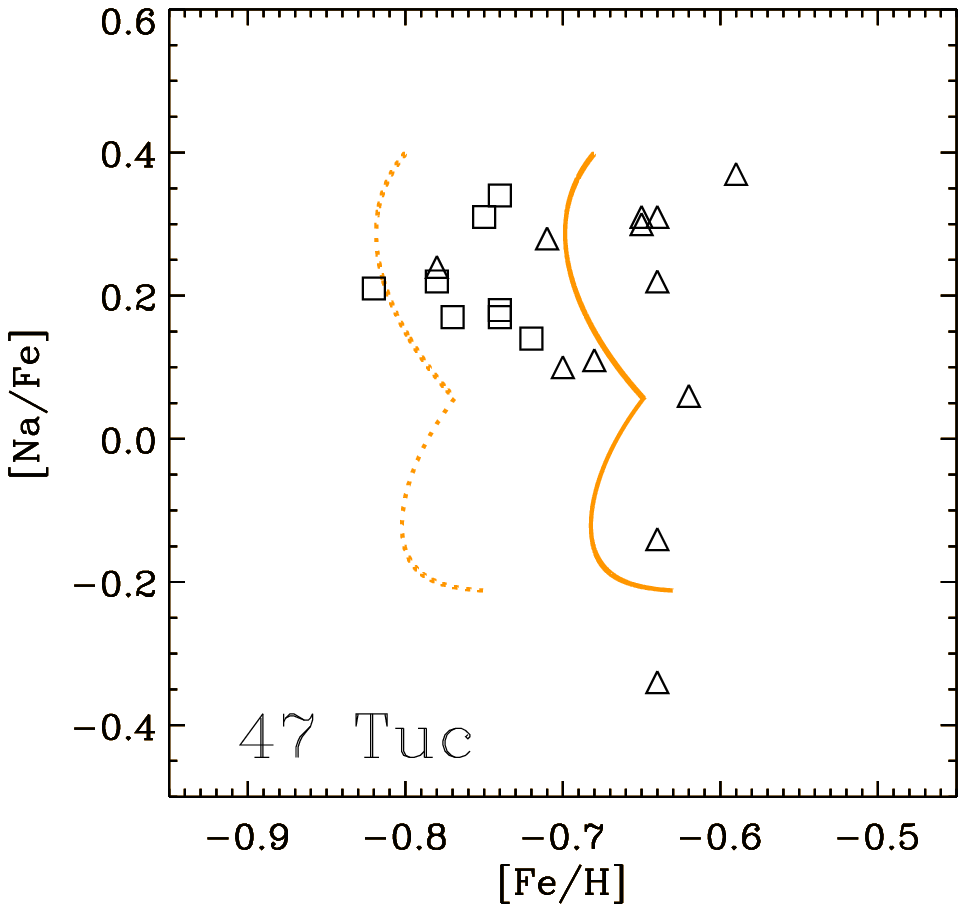,width=0.40\textwidth} 
\end{center}    
\caption{Left panel: evolution of [O/Fe] versus [Fe/H] for the model of 
globular cluster M~15 compared with the observational dataset from 
\citet{sneden1997}. Right panel: evolution of [Na/Fe] versus [Fe/H] for 
the model of globular cluster 47~Tuc compared with the observational 
datasets of \citet[][triangles]{carretta2004} and 
\citet[][squares]{koch2008}. For these very metal-poor and metal-rich 
GCs, the [Fe/H] remains roughly constant throughout their respective 
evolution.}
\label{fig:feo_M15}  
\end{figure*}

Figure~\ref{fig:feo_M15} shows that in agreement with the results of 
Paper~I for intermediate metallicity GCs, the [Fe/H] remains constant 
during the evolution of metal-poor {\it and } metal-rich globular 
clusters. (note that we choose to plot [Na/Fe] instead of the usual 
[O/Fe] because most of the oxygen values for this cluster are upper 
limits).

\section{M4 and its Fluorine content}

Fluorine abundances have been derived for the stars of M4 by Smith et 
al. (2005), finding they vary by a factor of six, correlate with oxygen, 
and anti-correlate with sodium and aluminum. Fluorine yields are a 
strong function of stellar mass and, therefore, can be used to constrain 
the nature of the polluters in globular clusters. Fluorine, like oxygen, 
is scarcely produced in intermediate-mass AGB stars (F is destroyed in 
5$-$6~M$_{\odot}$ and produced in the 4~M$_{\odot}$ models of Karakas \& 
Lattanzio 2007) and here, is mainly synthesised by the SNe~II models.

As is evident in the lower panels of Fig.~\ref{fig:M4}, where we plot 
the observational data alongside the predicted evolution of our model, 
fluorine correlates with oxygen and is anti-correlated with sodium. In 
our model, [F/Fe] evolves from an initially sub-solar value to a 
slightly super-solar value, typical of SNe~II pollution, and in 
agreement with the observational constraints. Smith et al. (2005) argue 
that the correlations can be explained within a self-polluting scenario 
in which AGB stars act as the main polluters. However, the destruction 
of fluorine by intermediate-mass AGB stars does not appear sufficient to 
explain the ``final'' extreme low values of the [F/Fe] ratio. In our 
model, these values are expected in the \it first \rm stars formed, due 
to the inhomogeneous SN~Ia effect.

\section{Other Elements}

\subsection{Heavy $\alpha$-elements} 
   
While there are multiple observations of light element abundances in 
GCs, only recently have heavier elements become accessible for 
statistically significant samples of stars. \citet{yong2005} measured 20 
elements in 38 bright giants of NGC~6752, while \citet{carretta2004} 
reported the abundances of nine sub-giants and three dwarfs of 47~Tuc, 
both for $\alpha$-elements and Fe-group elements. While light elements 
show the already discussed variations spanning in some cases more than 
an order of magnitude, silicon and heavier elements (X$_{\rm heavy}$) 
show much smaller star-to-star scatter in their [$X_{\rm heavy}$/Fe] 
ratio. This smaller scatter is consistent, in most cases, with 
measurement uncertainties.  Furthermore, the mean [$X_{\rm heavy}$/Fe] 
ratios are consistent with field stars of the same metallicity. There is 
some evidence for heavy $\alpha$- elements correlating with lighter 
elements; \citet{yong2008b} found a statistically significant, although 
loose, correlation between silicon and aluminum in NGC~6752, noting that 
the Si abundances are roughly constant from star to star.

\begin{figure*}     
\begin{center}     
\psfig{figure=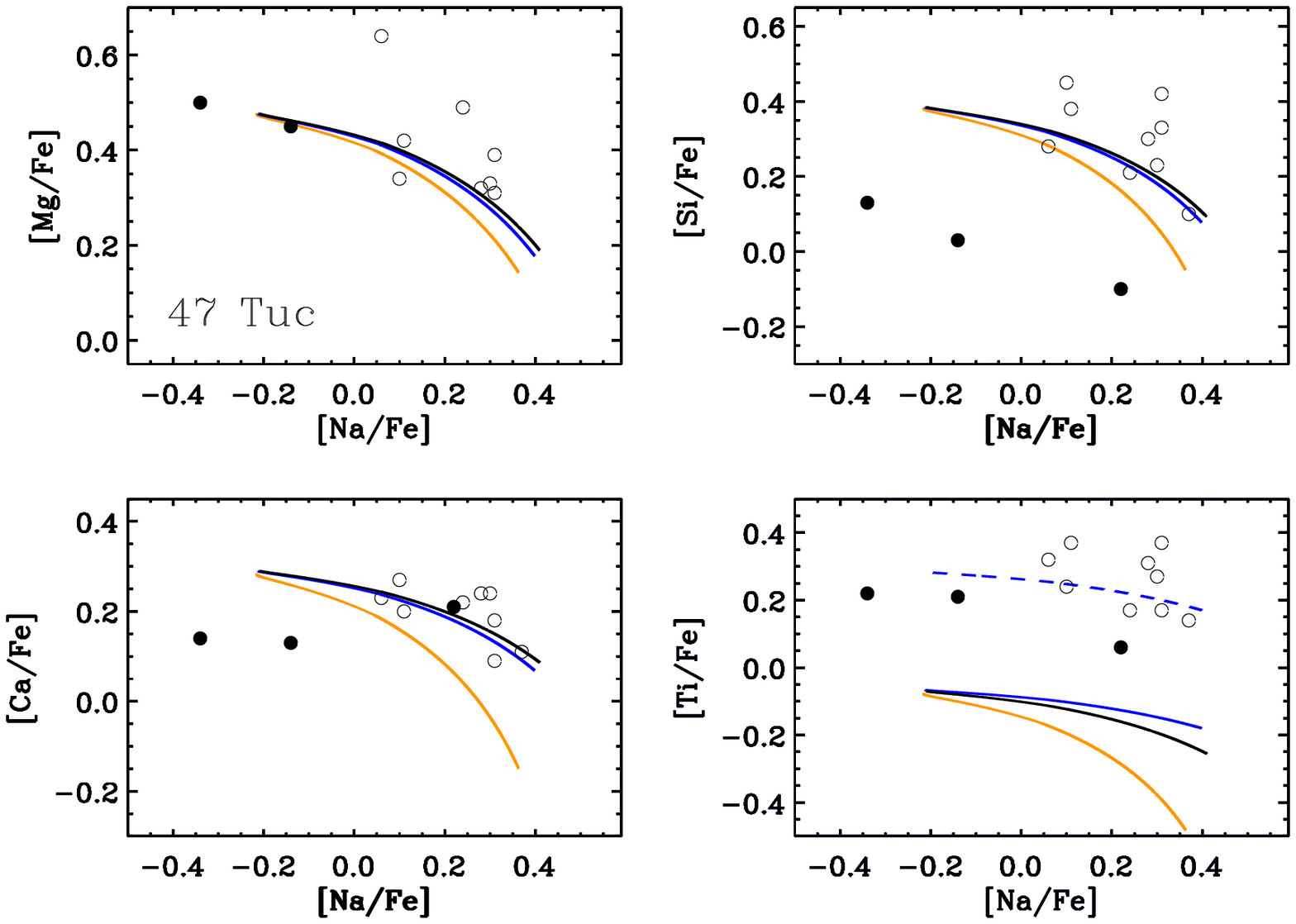,width=0.90\textwidth} 
\end{center}    
\caption{Evolution of different $\alpha$-elements ([Mg/Fe], [Si/Fe], 
[Ca/Fe] and [Ti/Fe]) versus [Na/Fe] for the case of 47~Tuc. Open circles 
correspond to the observational dataset from \citet{carretta2005}; the 
filled symbols are the three dwarfs, while the open symbols are the nine 
sub-giants. The colour-coded lines correspond to models with different 
SNe~Ia yields.  Orange lines: W7; blue lines: WDD1; black lines: CDD1. 
The dashed line in the bottom right panel corresponds to the [Ti/Fe] 
prediction using the WDD1 SNe~Ia model, but offset by $\sim$0.3~dex (see 
text for details).}
\label{fig:47Tuc_alpha} 
\end{figure*}

\begin{table}  
\centering  
\begin{minipage}{80mm}  
\caption{Isotope ratios for the averaged yields of SNe~II and SNe~Ia.
The labels are the same as used in Tables 1 and 2.} 
\label{tab:iso}  
\begin{tabular} {|l|c|c|c|}  
\hline  
             & $^{48}$Ti/$^{50}$Ti  &   $^{56}$Fe/$^{58}$Fe     & $^{58}$Ni/$^  {60}$Ni    
   \\ 
\hline 
SNe II (W\&W)   &      1.1e4             &      3.7e4                &     0.5      \\ 
SNe II (KOB)    &      2.3e2             &      1.7e3                &     0.36     \\  
SNe II (C\&L)   &      0.9e4             &      6.6e4                &     2.0      \\  
\hline 
Model  (SNe II) &       1.e4             &       5.e4                 &     0.5      \\ 
\hline   
SNe Ia (W7)      &      1.9               &      2.1e2                &     8.8       \\ 
SNe Ia (WDD1)    &      2.0               &      1.8e2                &     6.2       \\ 
SNe Ia (WDD3)    &      35.2              &      6.5e2                &     11.2      \\ 
\hline  
\end{tabular}    
\end{minipage} 
\end{table}

In Figure~\ref{fig:NGC6752_alpha} we plot the evolution of the [$X_{\rm 
i}$] ratio of different $\alpha$-elements versus the [O/Fe] for the 
model of NGC~6752, together with the observational values obtained by 
\citet{yong2005}. As can be seen, our model is able to reasonably 
reproduce the spread in the O-Mg correlation. The different lines in 
Figure~\ref{fig:NGC6752_alpha} represent models with different 
prescriptions for the SN Ia yields of \citet{iwamoto1999}: W7 (model 
with slow central deflagration), WDD1 (mode fast deflagration), and CDD1 
(model delayed detonation in the outer layer). As can be seen, the 
abundance of the Fe-group elements (in particular, the neutron-rich 
species such as $^{50}$Ti, $^{58}$Fe and $^{58}$Ni) depend considerably 
upon the chosen physics of the model. From Table~\ref{tab:yields}, it is 
evident that variations in the SNe~Ia yields by up to a factor of four 
are possible for these elements.

Intermediate-mass AGB stars produce some Mg - mostly in the form of the 
neutron-rich isotopes \citep{fenner2003} - and thus even if a variation 
of $\sim$1~dex is achieved in [O/Fe], only a mild depletion of 0.3 dex 
in [Mg/Fe] is present, due to the AGB contribution. The Mg production in 
different SN~Ia models is roughly the same, although it is small 
compared to the production of Mg in SNe~II and AGB stars.

An apparent failure of our model is that the ``shape'' of the O-Mg 
correlation is ``concave'', while it looks to be more ``convex'' in the 
observational data (upper left panel of Figure~6). This may be due to 
the fact that we are using averaged yields for SNe II. Indeed more 
massive SNe~II progenitors produce much larger amounts of Mg and explode 
earlier, explaining why the empirical [Mg/Fe] increases quite quickly, 
at apparent odds with our model.

While the contribution from the SN~Ia is not relevant for Mg, this is 
not the case for the other heavier $\alpha$-elements. Indeed, 
Table~\ref{tab:yields} shows that a SN~Ia produces more Si, Ca, and Ti 
than a single SN~II (note however that their production ratio against 
iron, [$X_{\rm i}$/Fe], is lower than for the case of SNe~II and is 
usually sub-solar). Moreover, variations in the initial value of the 
ratio [$X_{\rm i}$/Fe] (up to 0.2 dex for Si and 0.4 dex for Ca and Ti) 
are possible when testing different SNe~Ia models, even if all these 
values remain, more or less, sub-solar (see 
Figure~\ref{fig:NGC6752_alpha}). In addition, our models predict a 
correlation between Si-O and Ca-O, with variations of up to $\sim$0.5 
dex in the [Si/Fe] abundance and $\sim$0.3 dex for [Ca/Fe] (for models 
CDD1 and WDD1). These predicted correlations are not in agreement with 
the observational constraints. Adopting the W7 model increases the 
scatter, only making the situation worse.

For Ti, the models CDD1 and WDD1 maintain the [Ti/Fe] abundance at a 
roughly constant value, consistent with the observations (even if there 
is an intrinsic offset of $\sim$0.2~dex; dashed line in the bottom right 
panel of Figure~6). In contrast, model W7 produces [Ti/Fe] abundances 
that do not agree with the observations. Since Ca is not synthesized in 
intermediate-mass AGB stars, we investigate how much Ca production we 
would require in (very) metal-poor SNe~Ia to fit the observational 
constraints. Indeed, a factor of two more Ca production in (very) low 
metallicity SN~Ia \footnote{Or just the prompt SNe~Ia which explode on 
timescales shorter than $\sim$100~Myr.} compared with the WDD1 model 
(green line in Figure~\ref{fig:NGC6752_alpha}) is able to reconcile our 
model with observations. Note that the fact the [Ca/Fe] ratio in halo 
field stars is observed to decline with [Fe/H] implies that the Ca 
yields of \citet{iwamoto1999} are consistent with SN~Ia at larger 
metallicity.
 
The same experiment can be made for Si, but in this case we should also 
try to take into account a possible higher AGB contribution. As already 
discussed, \citet{yong2005} found a statistically significant 
correlation (at odds with our model) between the [Si/Fe] and [Al/Fe] 
which, in their interpretation, can be explained if the reaction 
$^{27}$Al(p,$\gamma$)$^{28}$Si is favoured over 
$^{27}$Al(p,$\alpha$)$^{24}$Mg. Hot-bottom burning (HBB) in 
intermediate-mass AGB stars is expected to produce small amounts of 
$^{28}$Si from proton capture on $^{27}$Al, though the Si yields are 
expected to be small \citep{karakas2003, karakas2007}. The Si production 
depends on the temperature of the HBB region and also on the assumed 
reaction rates. Again, similar to Ca, the problem of Si can be solved 
with a higher production of Si in (very) low metallicity SN~Ia 
\footnote{Or, again, prompt SN~Ia; as before, the value of 
[Si/Fe]$\sim$$+$0.0 in field stars at solar metallicity implies that at 
higher metallicity, the Iwamoto et~al. (1999) yields are correct (but 
note that these yields are calculated for solar metallicity).} and/or a 
higher Si production in AGB stars (a mean value of $3.5 \times 10^{-3}$ 
M$_{\odot}$ \citep[compared to $4.8 \times 10^{-5}$ M$_{\odot}$ 
predicted by][]{karakas08}). Note that this is {\it exactly} the value 
of Al we needed in Paper~I to reproduce the Al-Mg anti-correlation in 
this GC. {\it If} our framework is correct, it means that current 
intermediate-mass AGB models are underestimating the yields of Al and Si 
by 1-to-2 orders of magnitude.

From Figure~\ref{fig:47Tuc_alpha}, we see that are no serious issues in 
reproducing the same $\alpha$-elements for the more metal-rich cluster 
47~Tuc.\footnote{In Figure~\ref{fig:47Tuc_alpha}, we have shown the 
[$\alpha$/Fe] trends against [Na/Fe], rather than [O/Fe], as the 
majority of the stars in the \citet{carretta2005} sample either have 
non-detections or only upper limits to their oxygen abundances.} In this 
case, the collection of sub-giants and dwarfs from \citet{carretta2004} 
are consistent with our model. We should note that due to the higher 
SNe~II pre-enrichment ([Fe/H]$_{\rm ISM}$), this model does not reach 
the low values of [O/Fe] seen in the NGC~6752 model. In this case, the 
Iwamoto et~al. (1999) yields have no problem in fitting the 
observational dataset. In our framework, 47~Tuc should form later than 
NGC~6752 (i.e., from a more metal-rich ISM) and the suggestion that very 
low metallicity SN~Ia (or AGB) may have slightly different (up to a 
factor of four) Si and Ca yields is plausible.  In this case, the 
temporal evolution goes from high values of [Na/Fe] to low values.

\subsection{Fe-peak elements} 

SNe~Ia and SNe~II synthesise a significant fraction of the Fe-group 
elements, while intermediate-mass AGB stars contribute very little of 
the same. The analysis of [$X_{\rm i}$/Fe], where $X_{\rm i}$ represents 
an Fe-peak element, can prove extremely useful in understanding the 
relative contributios of the two types of supernovae. The ratio of iron 
produced in SNe~Ia-to-SNe~II is Fe$_{\rm SNe Ia}$/Fe$_{\rm SNeII}\simeq 
7-8$ (depending on the adopted yields); \it if \rm [$X_{\rm 
i}(SNeIa)$/$X_{\rm i}(SNII)$] $<$7$-$8, then [$X_{\rm i}$/Fe] will 
increase during the GC evolution and viceversa.

\begin{figure*}     
\begin{center}     
\psfig{figure=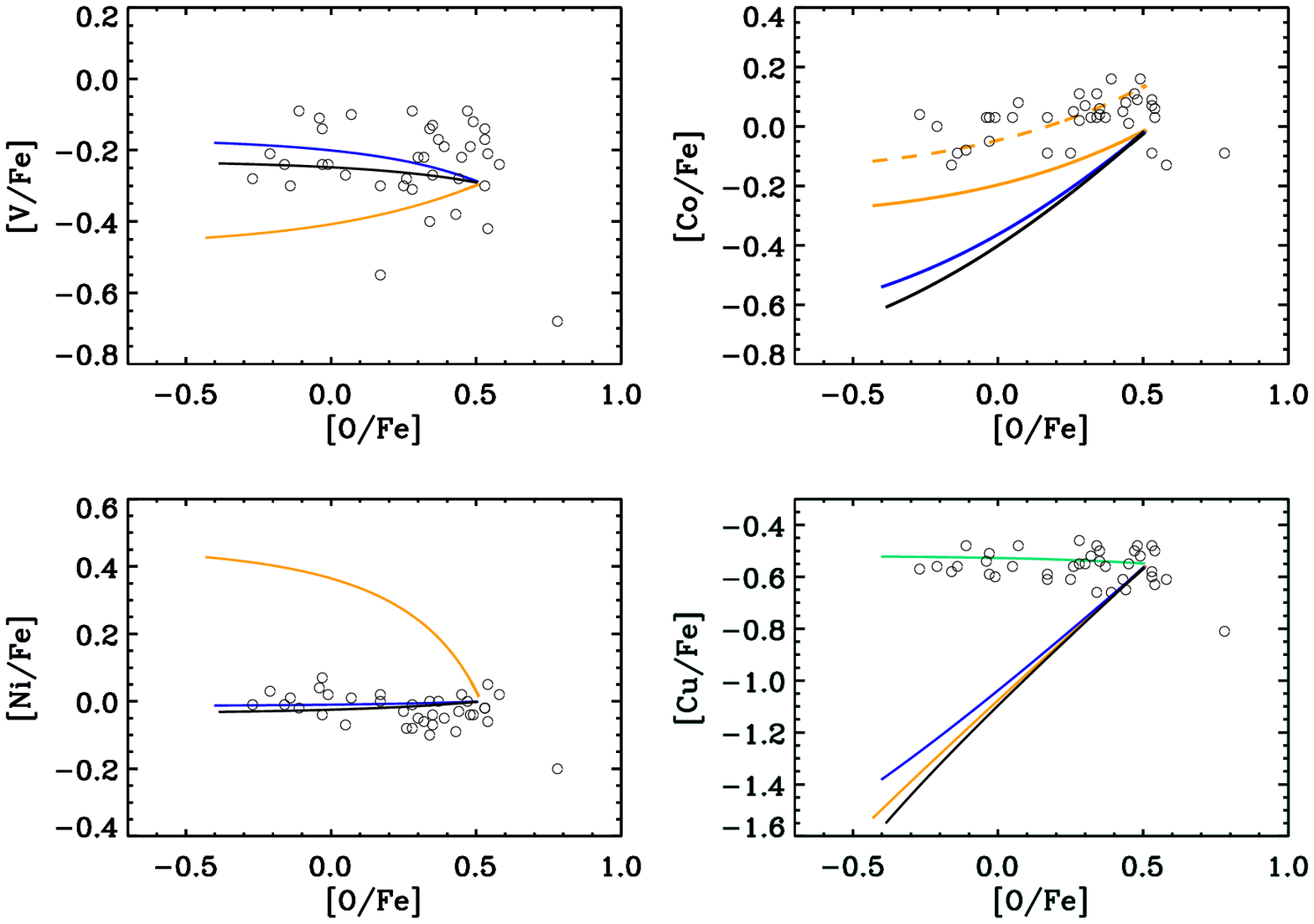,width=0.90\textwidth}  
\psfig{figure=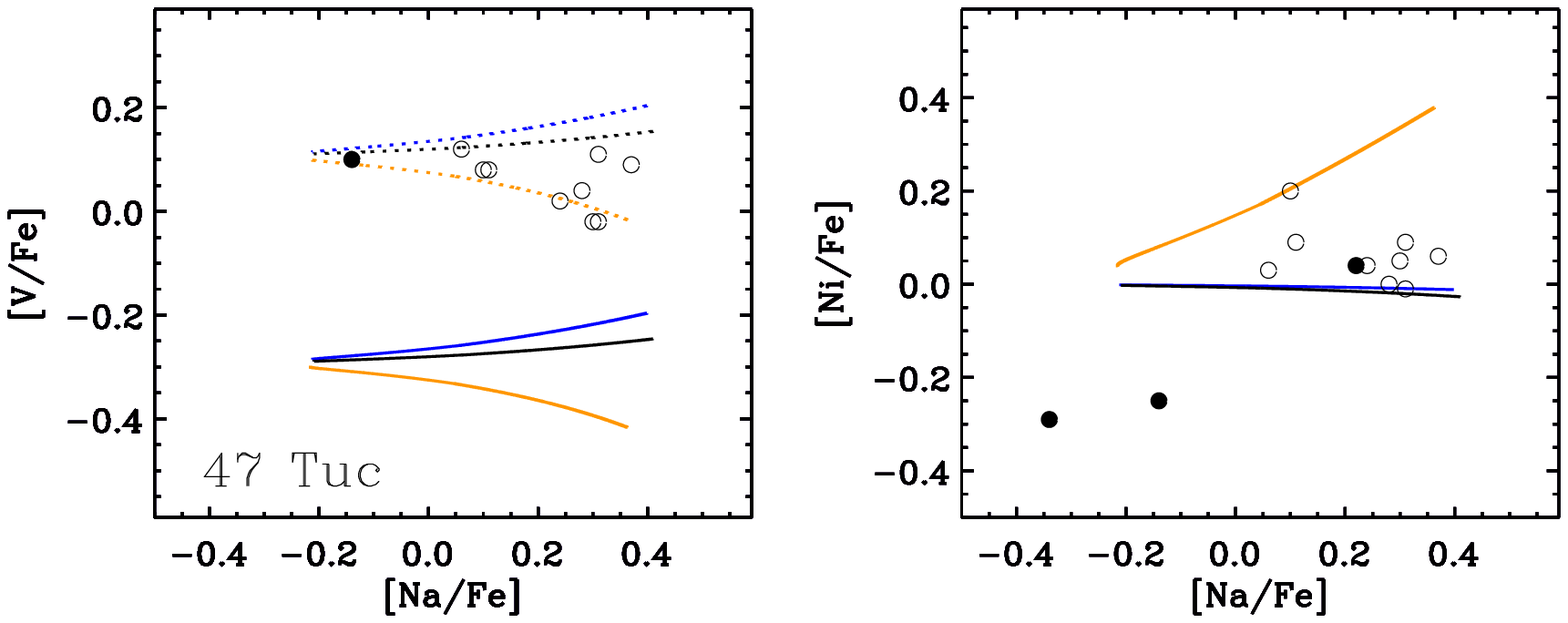,width=0.90\textwidth}   
\end{center}    
\caption{\it Upper Four Panels\rm: Evolution of Fe-group elements 
([V/Fe], [Co/Fe], [Ni/Fe], and [Cu/Fe]) versus [O/Fe] for the case of 
NGC~6752; open symbols correspond to the observational data from 
\citet{yong2005}. \it Bottom Two Panels\rm: Evolution of Fe-group 
elements ([V/Fe] and [Ni/Fe]) versus [Na/Fe] for the case of 47~Tuc; 
open symbols correspond to the observational data from 
\citet{carretta2005}. In all six panels, the colour-coded lines refer to 
models with different SNe~Ia yields. Orange lines: W7; blue lines: WDD1; 
black lines: CDD1. Dashed lines in the upper right and bottom left 
panels correspond to the arbitrary $\sim$0.1$-$0.3~dex model offsets 
required to match the extant data (see text for details). The green line 
in the middle-right panel corresponds to a model in which the Cu 
production in SNe~Ia is assumed to be $\sim1.2 \times 10^{-4}$ 
M$_{\odot}$, in order to match the observational constraints.}
\label{fig:NGC6752_fe-group}  
\end{figure*}

Figure~\ref{fig:NGC6752_fe-group} compares the results of our models for 
Fe-group elements (specifically V, Co, Ni, and Cu) with the 
corresponding observational data of \citet{yong2005}, for the case of 
the globular cluster NGC~6752. [V/Fe] and [Ni/Fe] remains roughly 
constant, in good agreement with observations, when using the WDD1 and 
CDD1 models. Using the W7 model for SN~Ia, though, produces less V and 
more Ni than needed to keep [Ni/Fe] and [V/Fe] constant, leading to 
variations of 0.2 and 0.4 dex for [V/Fe] and [Ni/Fe], respectively (see 
Table~\ref{tab:yields}). The trend in the evolution of Co is 
well-reproduced (apart from a small $\sim$0.1~dex offset) by using 
SNe~Ia model W7, while the others underestimate its production by 
roughly a factor of two.

In the lower two panels of Figure~\ref{fig:NGC6752_fe-group}, we compare 
the [V/Fe] and [Ni/Fe] evolution of our model for the metal-rich GC 
47~Tuc, with the data presented in \citet{carretta2005}. The models 
underproduce vanadium by a factor of two, and overproduce nickel by 
$\sim$0.2~dex.  We note in passing that observationally, the two 
clusters themselves differ in their mean [V/Fe] by $\sim$0.3~dex.  The 
origin of this difference remains unclear and in the lower left panel of 
Figure~\ref{fig:NGC6752_fe-group}, we show our model predictions for the 
47~Tuc [V/Fe]-[Na/Fe] evolutionary trend we show the same curves offset 
by $\sim$0.3~dex, to coincide with the data. We cautiously suggest that 
our model can (roughly) match the evolution of most of the Fe-group 
elements analysed, within the theoretical uncertainties of SNe yields, 
and those associated with the observational data.

Having just made that conclusion though, one notable exception exists, 
in the form of copper. Indeed, while the observed Cu content for 
NGC~6752 seems to be roughly constant, our model predicts a very strong 
correlation with oxygen, independent of the chosen SNe~Ia model. From 
the middle-right panel of Figure~\ref{fig:NGC6752_fe-group}, we see that 
the initial [Cu/Fe] ratio is greatly underestimated in our models, for 
any choice of the SNe~Ia yields: all the SNe~Ia models analysed here 
produce very little copper (see Table~\ref{tab:yields}). The origin of 
the copper observed in field halo stars remains a matter of debate and, 
at this stage, the magnitude of this apparent failure of the model needs 
to be confirmed/refuted with additional empirical data. \it Assuming \rm 
a value for SNe~Ia copper production of $1.2\times10^{-4}$ M$_{\odot}$ 
would bring the models into agreement with the extant data. It is very 
interesting to note that such a suggestion is not entirely without 
precedent; indeed, this value is very similar to the one 
($0.5-2.0\times10^{-4}$ M$_{\odot}$) proposed by \citet{matteucci1993} 
to reproduce the [Cu/Fe] versus [Fe/H] trend for Milky Way stars 
\citep[but see also][]{romano2007b}.  In addition, 
\citet{mishenina2002}, analysing the copper abundance trend in 90 
metal-poor stars in the metallicity range $-3.0 \ge$[Fe/H]$\ge -0.5$, 
found evidence that SNe Ia must play a significant role ($>$65\%) in 
producing copper.
   
\citet{yong2008b} pointed out that even if the abundance of Fe-peak 
elements shows a small scatter in NGC~6752, their values statistically 
correlate with nitrogen. This is exactly what we expect in our model, 
even if the exact slope and scatter of this correlation depends upon the 
interplay between a SN~Ia origin and a SN~II elemental origin.

\section{Isotope ratios as an observational test to validate the framework} 
 
In Paper~I, we tested our model predictions for the behaviour of Mg 
isotopes in GCs against measured isotope abundances in NGC~6752 and 
M~13. Our framework succeeded in reproducing their observed trends, but 
we require a better test to verify the SN~Ia framework of our model 
(recall, Mg is not a strong constraint for the SN~Ia framework, as its 
origin is linked to SNe~II and AGB stars). The purpose of this section 
is to make predictions regarding isotope ratio of heavy elements where 
the imprint of a SN~Ia should be more important; this can constitute a 
definitive test for the model.

As already noted by \citet{iwamoto1999}, even if different SN~Ia models 
can produce slightly different yields, they all agree that a SN~Ia 
should produce heavy neutron-rich isotopes, such as $^{50}$Ti, 
$^{58}$Fe, and $^{58}$Ni. In Table~\ref{tab:iso} we summarise the 
isotope ratios of these elements for different SNe~II and SNe~Ia models 
in the literature. For the following, the SNe~II yields we use are 
labeled ``SNe II model'' in Table~\ref{tab:iso}, while we will test the 
different ratios associated with different yields for SNe~Ia.
  
In Figure~10, we show the predictions of our NGC~6752 and 47~Tuc 
models for the 
evolution of the isotope ratios $^{48}$Ti/$^{50}$Ti, 
$^{56}$Fe/$^{58}$Fe, and $^{58}$Ni/$^{60}$Ni versus [O/Fe]. As can be 
seen, variations of up to two orders of magnitude are predicted and 
should be observed if inhomogeneous pollution by the SN~Ia is the 
condition for a GCs formation. Indeed as $^{50}$Ti and $^{58}$Fe are 
primarily produced in SN~Ia, and marginally produced in SNe~II, the 
$^{56}$Fe/$^{58}$Fe and $^{48}$Ti/$^{50}$Ti isotope ratios increase 
considerably during the formation of the GC. The case of $^{58}$Ni and 
$^{60}$Ni is a bit different since both isotopes are produced by both 
types of SNe. In this case, the $^{58}$Ni production by SN~Ia should be 
observable even if there is $only$ a change of 0.8-1.0 dex in the 
$^{58}$Ni/$^{60}$Ni ratio, which decreases during the evolution.

While the measurements of such isotope ratios would help to test our 
framework, from an observational perspective such measurements are 
likely beyond the limit of current instrumentation. It might be possible 
to measure Fe isotope ratios from the FeH molecular lines; such lines 
are most likely to be present in extremely cool dwarfs.
  This is 
beyond the observational capabilities of current instrumentation but 
it might be possible using projected high-resolution near-infrared 
spectrographs on 40m-class telescopes.
Isotope ratios for Ti have been measured in 
only a handful of near-solar metallicity stars \citep[e.g.,][] 
{lambert1977, chavez2009}. These isotopes can only be measured from TiO 
molecular lines, and such lines are only present in metal-rich giants 
([Fe/H] $\ge -0.7$) or very cool dwarfs. Since NGC~6752 is reasonably 
metal-poor, the only possibility of finding TiO lines would be at the 
very cool end of the main sequence, beyond current instrumentation. 
Owing to its higher metallicity, 47~Tuc ([Fe/H]$\sim0.7$) may be an 
excellent candidate to test our predictions with current capabilities
while measurements in other, more metal poor GC will be possible with 
the advent of the 40-m class telescopes.

As shown by \citet{hughes2008}, the \citet{chavez2009} data suggest that 
mildly sub-solar metallicity giants in the disc+halo show 
log($^{48}$Ti/$^{50}$Ti)$\approx$1.1, with very little variation from 
[Fe/H]$\approx$$-$0.7 to [Fe/H]$\approx$$+$0.0.  If confirmed in a 
metal-rich GC like 47~Tuc, this would suggest that the W7 or WDD1 SNe~Ia 
models are to be preferred over the CDD1 model (which is $\sim$0.7~dex 
higher in log($^{48}$Ti/$^{50}$Ti) in this metallicity range).

\begin{figure*}     
\begin{center}     
\psfig{figure=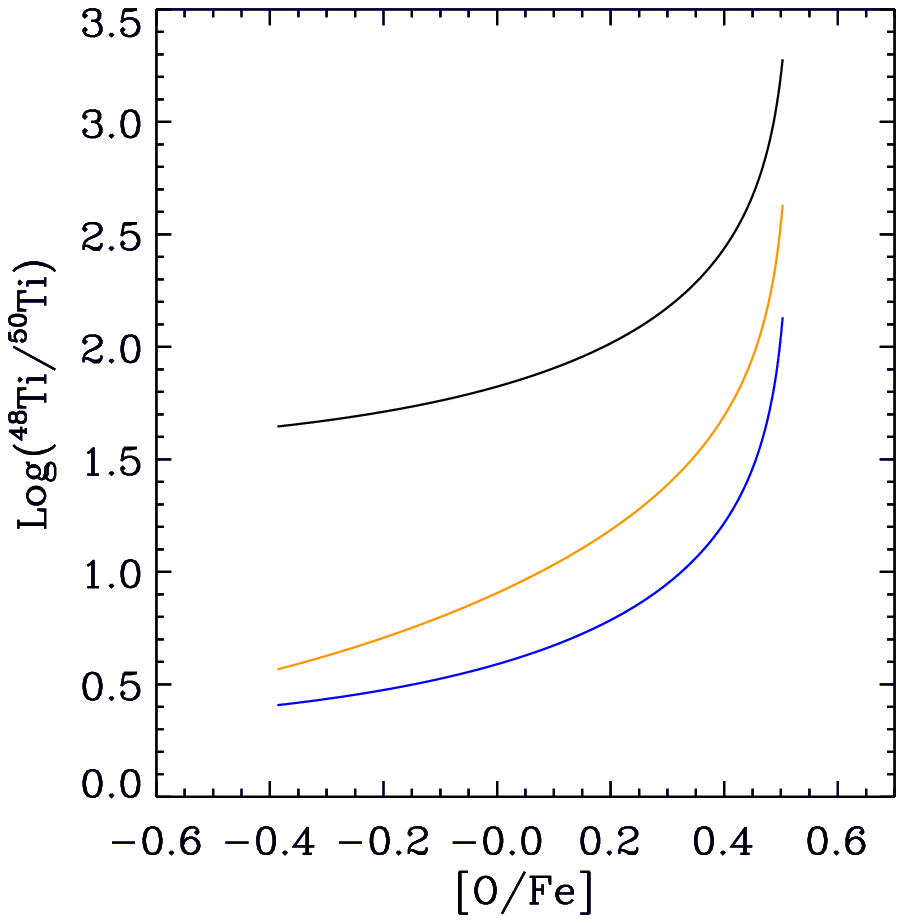,width=0.30\textwidth} 
\psfig{figure=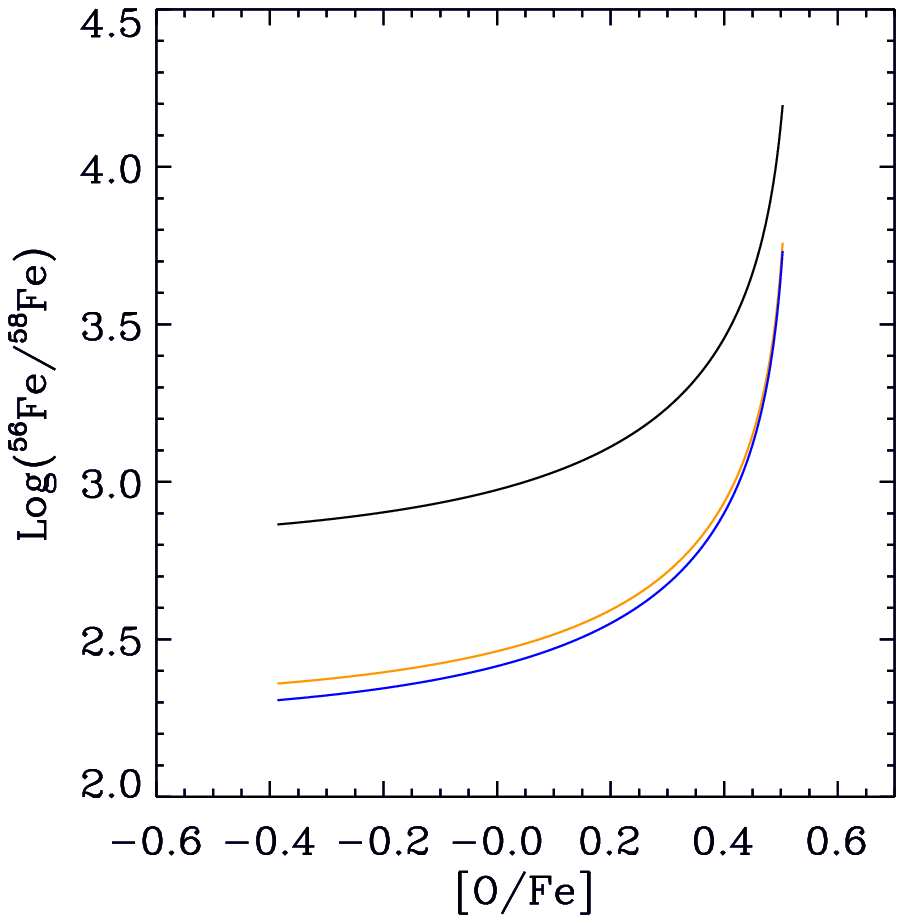,width=0.30\textwidth} 
\psfig{figure=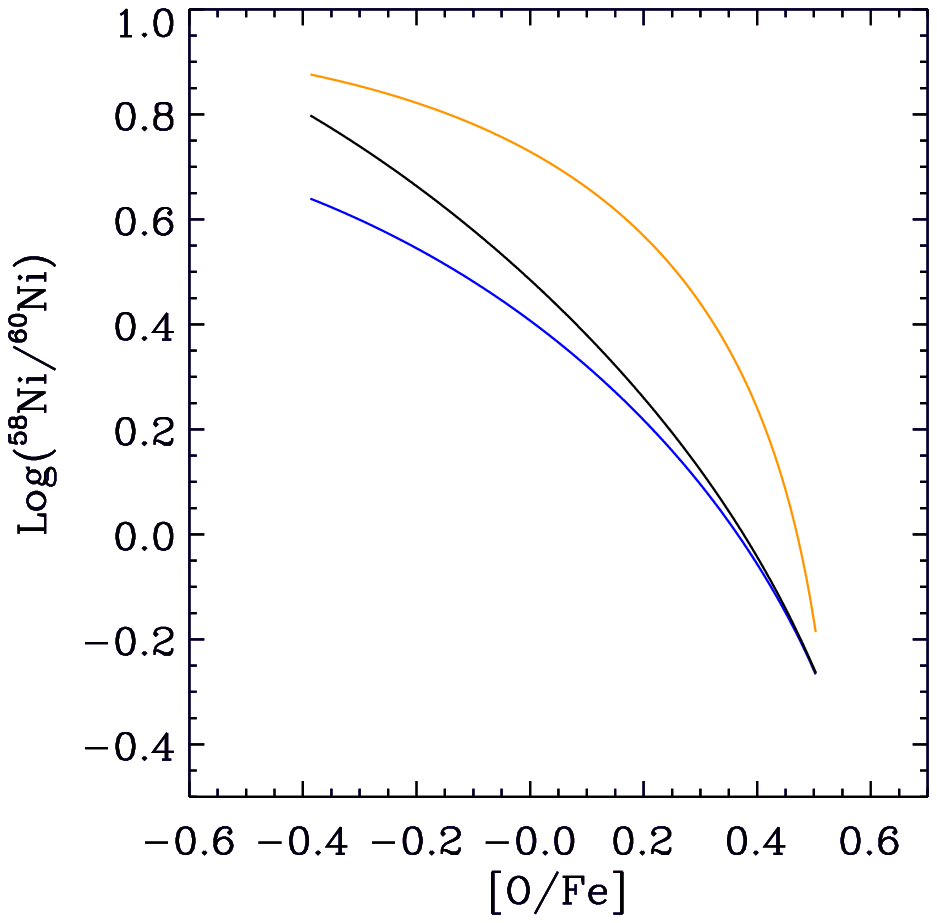,width=0.30\textwidth}
\psfig{figure=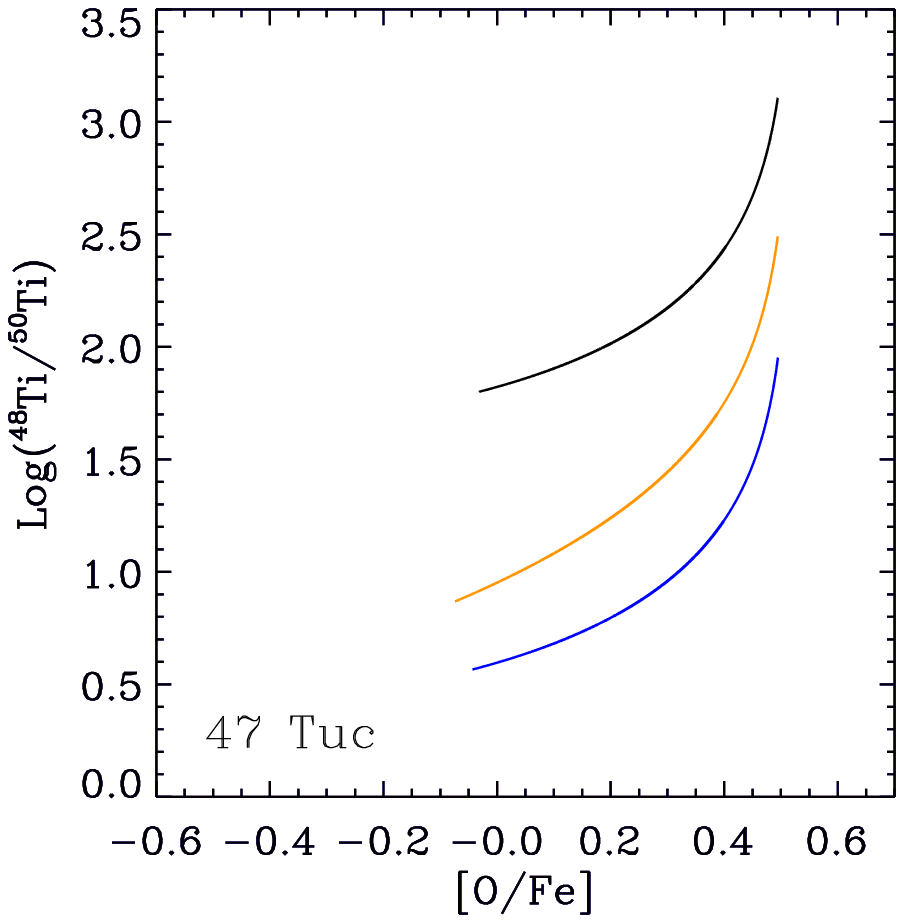,width=0.30\textwidth}
\psfig{figure=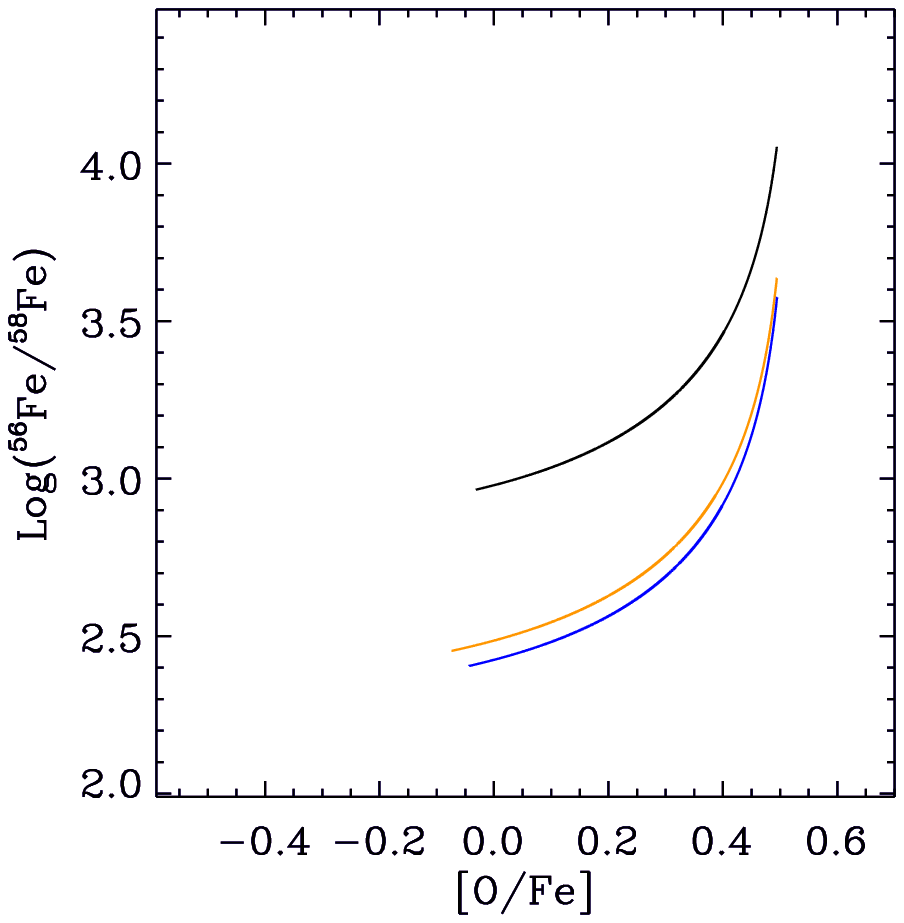,width=0.30\textwidth}
\psfig{figure=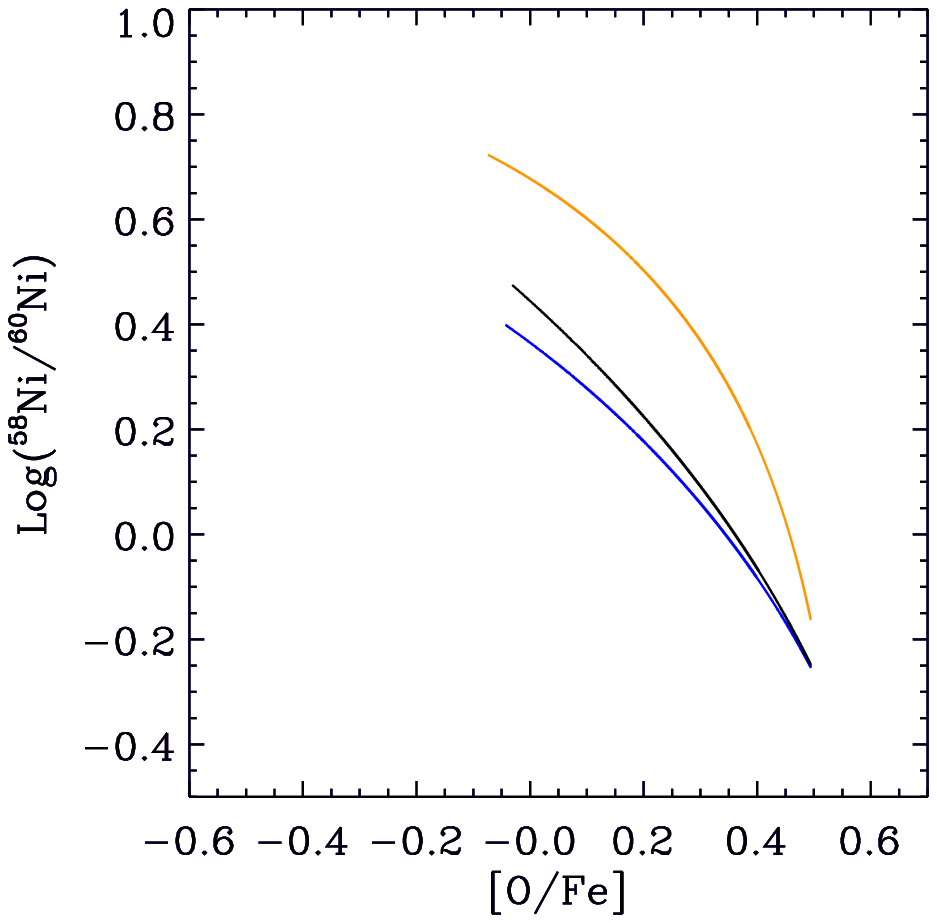,width=0.30\textwidth}
\end{center}    
\label{fig:iso} 
\caption{Evolution of the isotope ratios for $^{48}$Ti/$^{50}$Ti, 
$^{56}$Fe/$^{58}$Fe, and $^{58}$Ni/$^{60}$Ni versus [O/Fe], for the 
NGC~6752 model (upper panels) and for 47~Tuc (lower panels). 
The isotopes $^{50}$Ti, $^{58}$Fe, and $^{58}$Ni are 
mainly produced by SNe~Ia \citep{iwamoto1999}.  The colour-coded lines 
correspond to models with different SNe~Ia yields.  Orange lines: W7; 
blue lines: WDD1; black lines: CDD1.}
\end{figure*} 
 
\section{Conclusions} 
 
In this paper we continue the study of the chemical evolution of GCs in 
the new framework of ``peculiar pre-enrichment'' presented by 
\citet{marcolini2008}. We extend the previous study to the very 
metal-poor globular clusters M~15 and NGC~6397 and the metal-rich 
cluster 47~Tuc. We also study the chemical evolution of fluorine in M~4. 
Our major findings can be summarised as follow:
 
\begin{enumerate} 
 
\item In our framework we can reproduce the sodium-fluorine 
anti-correlation, as well as the oxygen-fluorine correlation observed in 
M~4.
 
\item The model can reproduce the very low [Fe/H] contents of M~15 and 
NGC~6397 assuming that they formed early in the history of the halo, and 
that the SN~Ia+AGB stars' inhomogeneous pollution was localised in a 
larger radius (compared to the more metal-rich clusters). The Na-O and 
C-N anti-correlations are also well-reproduced for these cases.

\item The light-element anti-correlations are also reproduced in the 
case of the metal-rich globular cluster 47~Tuc, which in our framework 
should form at a later stage of the halo's formation, with the 
inhomogeneous pollution confined to a much smaller radius (e.g., 
$\sim$20~pc).
 
\item We also tested our model against the observed $\alpha$-element 
abundances (O, Mg, Si, Ca, and Ti) for different SNe~Ia models. In 
general, the slow deflagration SN~Ia models (WDD1 and CDD1) of 
\citet{iwamoto1999} better reproduce the observational constraints than 
the fast deflagration (W7) models. This is because the WDD1 and CDD1 
models produce more Si, Ca, and Ti. The model reproduces the 
$\alpha$-element content in the case of the metal-rich globular 47~Tuc, 
where a slight Na-Si and Na-Ca anti-correlation is observed. In the case 
of the intermediate-metallicity globular NGC~6752 we predict a moderate 
Si-O and Ca-O correlation which is not compatible with observations. 
However, this problem can be solved by assuming that (very) low 
metallicity (or prompt) SNe~Ia produce a factor of four (two) more Si 
(Ca) than the yields prescribed by \citet{iwamoto1999}, that were 
computed for solar metallicity. Also, more Si production by 
intermediate-mass AGB stars would also help to solve this problem. This 
last point, though, is not compatible with current AGB models.
 
\item The model predicts that the evolution of most of the Fe-peak 
elements should remain approximately constant (V and Ni), or slightly 
anti-correlated with oxygen (Co). This is because SNe~Ia yields 
prescribe a large production of the Fe-group elements (with the slow 
deflagration model better fitting the observations). However, while our 
model in general predicts that Fe-group elements should slightly 
correlate or anti-correlate with lighter elements, the details (and the 
scatter) depends strongly on the adopted SNe~II yields, with yields at 
different metallicities required (especially for SNe~Ia) for a more 
detailed analysis.
 
\item The only notable exception is copper, which we fail to reproduce 
by more than an order of magnitude when using yields from the 
literature.  This is due to the low Cu production in theoretical SNe~Ia 
models.  We propose a value for the copper production of $\sim1.2 \times 
10^{-4}$ M$_{\odot}$ in SNe~Ia, to match the observational constraints. 
While copper production is still a matter of debate in the literature, 
it is worth noting that our proposed value is essentially the same 
suggested by \citet{matteucci1993}, to reproduce the copper abundances 
of the disk+halo of the Milky Way.
 
\item Finally, we propose that the discovery of high isotopic ratios (at 
high [$\alpha$/Fe]) involving the neutron-rich $^{50}$Ti, $^{58}$Fe, and 
$^{58}$Ni species could be used as an observational test of our model. 
This is because these are mainly produced by SN~Ia, unlike the Mg 
isotopes employed in Paper~I which are produced by both SNe~II and AGB 
stars. This test will likely only be possible for the most metal-rich 
clusters, such as 47~Tuc.

\item  The dynamical feasibility of the proposed scenario, 
however, remains to be probed wiht hydrodynamical simulations.

\end{enumerate}

\section*{Acknowledgments} 
 
We kindly thank David Yong for his insights into isotopic abundance 
determinations.  PSB is supported by the Ministerio de Ciencia e 
Innovaci\'on (MICINN) of Spain through the Ramon y Cajal programme.
PSB also acknowledges a Marie Curie Intra-European Reintegration grant 
within the 6th European framework program {\it and} financial support from the 
Spanish Plan Nacional del Espacio del Ministerio de 
Educaci\'on y Ciencia (AYA2007-67752-C03-01).
BKG acknowledges the support of the UK's Science \& 
Technology Facilities Council (STFC: ST/F002432/1, ST/G003025/1), 
the UK's National Cosmology 
Supercomputer (COSMOS), the University of Central Lancashire's High 
Performance Computing Facility, and the generous financial 
support of Saint Mary's and Monash Universities visitor programs. 
AIK acknowledges support from the Australian Research Council's 
Discovery Projects funding scheme (DP0664105). 

\bibliography{globular_refs}     
    
\label{lastpage}     
\end{document}